\documentclass[aps,pra,twocolumn,a4paper]{revtex4-1}
\usepackage{newtxtext,newtxmath,braket}
\usepackage[pdftex]{graphicx,color}

\begin{document}
\title{
Efficient detection of inhomogeneous magnetic fields from a single spin with Dicke states 
}
\author{Hideaki Hakoshima}\email{hakoshima-hideaki@aist.go.jp} \author{Yuichiro Matsuzaki}\email{matsuzaki.yuichiro@aist.go.jp} %
\affiliation{
Nanoelectronics Research Institute,
National Institute of Advanced Industrial Science and Technology (AIST)\\
1-1-1 Umezono, Tsukuba, Ibaraki 305-8568, Japan
}
\date{\today}
\begin{abstract}
The efficient detection of a single spin is a significant goal of improving the sensitivity of quantum magnetic-field sensors.
Recent results show that a specific type of entanglement such as Greenberger-Horne-Zeilinger (GHZ) states
can be used as a resource to improve the performance of single spin detection.
However, scalable generation of the GHZ states is experimentally difficult to realize.
It is desirable to use a practical entangled state that can be easily generated.
In this paper, we propose the efficient detection of a single spin with Dicke states.  
We show a way to prepare and measure Dicke states via a global control.
Moreover, we investigate how dephasing due to unwanted coupling with the environment affects the performance of our proposal,
and show that single spin detection with Dicke states with dephasing
has a significant advantage over the classical strategy with separable 
states. Our results are important toward realizing entanglement enhanced single spin detection.
\end{abstract}
\maketitle

\section{Introduction}
A great deal of effort has long been devoted to improving the accuracy of the measurement for weak magnetic fields and many types of magnetic sensors have been developed so far \cite{degen2017quantum}.
The precise measurement is not only fundamentally interesting (as it is related to  exploring the ultimate precision allowed by quantum mechanics) but also is important for practical applications in various fields of study such as condensed-matter physics, material science, and life sciences \cite{schirhagl2014nitrogen}.
Particularly, the efficient detection of a single (electron or nuclear) spin \cite{degen2008nanoscale,taylor2008high,maze2008nanoscale,balasubramanian2008nanoscale,schaffry2011proposed,muller2014nuclear,staudacher2013nuclear,mamin2013nanoscale,ohashi2013negatively,rugar2015proton,lovchinsky2016nuclear,zhao2012sensing,shi2015single,abe2018tutorial,rugar2004single,cujia2019tracking} is an extremely important task and also one of the ultimate goals in quantum metrology.
However, the magnetic field from the single spin is weak, and a long total measuring time is required to detect with the current technology. Therefore, it is essential to improve the sensitivity of the magnetic field for more rapid detection of the single spin.

It is known that entanglement can be a resource to achieve sensitivity for homogeneous magnetic fields beyond the standard quantum limit 
(SQL) 
\cite{leibfried2004toward,giovannetti2004quantum,pezze2008mach,giovannetti2011advances,jones2009magnetic,matsuzaki2011magnetic,chin2012quantum,macieszczak2015zeno,tanaka2015proposed,dooley2016quantum,pezze2018quantum,matsuzaki2018quantum,tatsuta2019quantum}.
Especially, the Greenberger-Horne-Zeilinger (GHZ) state (which is also called a cat state) achieves the highest sensitivity without any noise, which is determined by the Heisenberg uncertainty relation, the so-called Heisenberg limit.
Recently, it was shown that the GHZ states can overcome the SQL even under 
the effect of realistic decoherence \cite{matsuzaki2011magnetic,chin2012quantum}.
In addition to the case of homogeneous magnetic fields, the GHZ states are also useful for the detection of the inhomogeneous magnetic fields from a single spin \cite{hakoshima2020single}.
Due to the dipole-dipole interaction between the target single spin and probe spins, the magnetic fields are inversely proportional to the distance cube from the target single spin, and therefore the magnetic fields affected by probe spins are quite different from the homogeneous magnetic fields.
Despite this great difference, the GHZ states can also detect single spin efficiently.  

However, it is known that the accurate control and measurement of the GHZ state is experimentally difficult to be realized, because it typically
requires accurate individual control of the qubits \cite{lu2014experimental,song2019generation,omran2019generation}.
To achieve a high sensitivity much better than the SQL, a large entangled state is necessary, which might be difficult to obtain as long as we use the GHZ states. Toward realizing
practical entanglement-based quantum sensors, it is essential to use an entangled state that can be measured just by a global control.

Here, we propose single spin detection by using Dicke states \cite{dicke1954coherence,campbell2009characterizing,rehler1971superradiance,skribanowitz1973observation,gross1982superradiance,angerer2018superradiant,lambert2016superradiance,tatsuta2018conversion,holland1993interferometric,kim1998influence,raghavan2001generation,toth2012multipartite,zhang2014quantum,apellaniz2015detecting,holland1993interferometric,kim1998influence,raghavan2001generation,andre2002atom,toth2012multipartite,zhang2014quantum,apellaniz2015detecting,wieczorek2009experimental,hume2009preparation,noguchi2012generation,ivanov2013creation,lamata2013deterministic,stockton2004deterministic,xiao2007generation,shao2010deterministic,wu2017generation,kasture2018scalable,ionicioiu2008generalized,shao2010deterministic,chakraborty2014efficient,bartschi2019deterministic} that can be created and measured by a global and deterministic control. 
Dicke states are related to a well-known cooperative phenomenon, ``phenomenon of superradiance'' that has been discussed for a long time
\cite{rehler1971superradiance,skribanowitz1973observation,gross1982superradiance,angerer2018superradiant,lambert2016superradiance}.
Dicke states are also highly entangled states, known to be a resource to measure spatially homogeneous magnetic fields with an accuracy beyond the SQL without decoherence \cite{holland1993interferometric,kim1998influence,raghavan2001generation,toth2012multipartite,zhang2014quantum,apellaniz2015detecting}.
In this paper, we show that Dicke states are also useful resources to detect  inhomogeneous magnetic fields from the single spin 
even under the effect of dephasing on the probe spins.
Although Dicke states are created in various methods of previous experiments \cite{wieczorek2009experimental,hume2009preparation,noguchi2012generation,ivanov2013creation,lamata2013deterministic} or theories \cite{stockton2004deterministic,xiao2007generation,shao2010deterministic,ionicioiu2008generalized,wu2017generation,chakraborty2014efficient,kasture2018scalable} such as by continuous measurement \cite{stockton2004deterministic} and quantum algorithms implemented as a quantum circuit \cite{ionicioiu2008generalized,chakraborty2014efficient,bartschi2019deterministic}, 
we propose a scheme to create and measure Dicke states by a global and deterministic control. 
To implement our protocol for spin detection, the necessary number of operational steps are constant against the size of Dicke states, and so our scheme is efficient in terms of scalability.

The remainder of this paper is organized as follows.
In section II, we describe the set-up of our scheme for single spin detection with Dicke states.
In section III, we show our results about the sensitivity of single spin detection with Dicke states
under the effect of dephasing 
In section IV, we describe our proposal to create and measure Dicke states via a global control. 
Finally, we summarize and conclude our paper in section V.

\section{Single spin detection with Dicke states}
In this section, we explain the details of our proposal for single spin detection with Dicke states.
Especially, we describe
the Hamiltonian, decoherence model, and our measurement basis. 
\begin{figure}
\centering
  \includegraphics[clip,width=8.5cm]{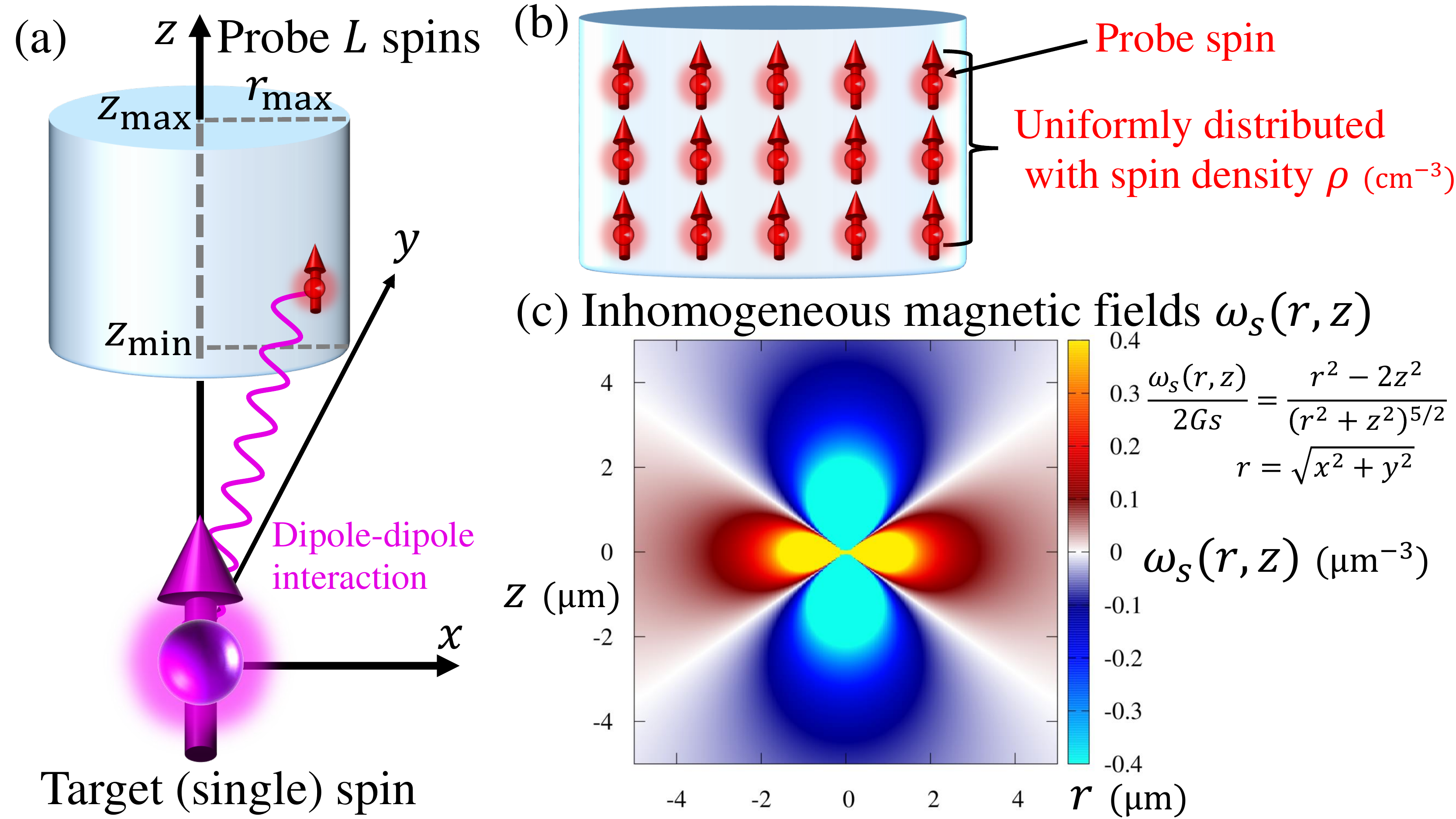}
  \caption{(color online) 
(a) We set the three-dimensional coordinate system.  
The target spin is located at $(0,0,0)$, and the probe $L$ spins are inside a columnar substrate, the center axis of which coincides with the $z$ axis.
The quantization axes of both the target spin and the probe spins are along the $z$ axis.
The quantization axes are determined by homogeneous external magnetic fields characterized by $\omega^{({\rm P})}$ and $\omega^{({\rm T})}$.
(b) The probe spins are uniformly distributed with the spin density of $\rho$ $ ($cm$^{-3})$.
(c) Plot of the magnetic-field strength $\omega_s(r,z)$ $ (\mu$m$^{-3})$ against $r$ $(\mu$m$)$ and $z$ $(\mu$m$)$  from the target spin.
Due to the dipole-dipole interaction between the target spin and the probe spins, the probe spins are affected by the inhomogeneous magnetic fields $\omega_s(r,z)$.
Through these magnetic fields, we estimate the state of the target spin to be up or down.}
 \end{figure}
We consider a single target spin and an ensemble of probe $L$ spins. 
For simplicity, we assume $L$ is an even number throughout the paper.
As shown in Fig.\ 1 (a) and (b), the target spin is located in the origin of the coordinate, and the probe spins are uniformly distributed inside a columnar substrate with the spin density of $\rho$.
Each probe spin is located at $\vec{r}_j=(x_j,y_j,z_j)$.
Since there is the dipole-dipole interaction between the target
spin and the probe spins, the Hamiltonian of the total system is given by
\begin{align}
\hat{H}&=\hat{H}_{{\rm T}}+\hat{H}_{{\rm P}}+\hat{H}_{{\rm I}},\\
\hat{H}_{{\rm T}}&=\frac{\omega^{({\rm T})} }{2}\hat{\sigma}_z^{({\rm T})},\quad \hat{H}_{{\rm P}}=\sum_{j=1}^L\frac{\omega^{({\rm P})}}{2}\hat{\sigma}_{z,j}^{({\rm P})},\\
\hat{H}_{{\rm I}}&=G\sum_{j=1}^L\frac{\vec{\hat{\sigma}}^{({\rm T})} \cdot \vec{\hat{\sigma}}_{j}^{({\rm P})}-3\left(\vec{\hat{\sigma}}^{({\rm T})} \cdot \frac{\vec{r}_j}{|\vec{r}_j|}\right)\left(\vec{\hat{\sigma}}_{j}^{({\rm P})}\cdot \frac{\vec{r}_j}{|\vec{r}_j|}\right)}{|\vec{r}_j|^3},
\label{eq:Hamiltonian}
\end{align}
where $\omega^{({\rm T})}$ ($\omega^{({\rm P})}$) is the Zeeman energy of the target (probe) spin, 
 $G=\frac{\mu_0\gamma^{({\rm T})}\gamma^{({\rm P})}}{16\pi}$ (here, we choose $\hbar=1$) is a constant determined by the magnetic moments of the target spin $\gamma^{({\rm T})}$ and the probe spins $\gamma^{({\rm P})}$,  
$\vec{\hat{\sigma}}_{j}^{({\rm P})}=(\hat{\sigma}_{x,j}^{({\rm P})},\hat{\sigma}_{y,j}^{({\rm P})},\hat{\sigma}_{z,j}^{({\rm P})})$ is a 
set of the Pauli matrices of the probe spins at $\vec{r}_j=(x_j,y_j,z_j)$, and $\cdot$ expresses the inner product. 
Here, we assume 
a large detuning between the target spin and the probe spins 
$\omega^{({\rm P})}\gg \omega^{({\rm T})}$.
This is a valid assumption when the probe spins are 
nitrogen vacancy (NV) centers in diamond and the target is a spin $1/2$ 
where the NV centers in diamond have a large zero-field splitting \cite{schirhagl2014nitrogen}.
In our paper, we assume that the effect of the dipole-dipole interaction between the probe spins is included in the dephasing term described by $T_2^*$ 
(a mathematical definition of which we will show below). Such an effect of $T_2^*$ is interpreted as 
a variation of the frequency of the probe spins, which suppresses 
a longitudinal relaxation due to the dipole-dipole interactions 
between probe spins. 
Actually, there are many experiments with high-density NV center ensembles, 
and theoretical models to treat the dipole-dipole interaction between NV centers 
as dephasing effect can reproduce the experimental results \cite{bauch2019decoherence,hayashi2020experimental,matsuzaki2016optically,kubo2011hybrid,zhu2014observation}. 
It is worth mentioning that, even if the inverse of $T_2^*$ is larger than 
the coupling from the target spin, we can in principle detect the target spin by
increasing the number of the repetitions of the measurements as follows.
First, we can perform a calibration measurement to know the effect of $T_2^*$
without the target spin. Next, we perform the actual measurement to detect the target spin.
The subtraction of the signals between the calibration and actual experiments
allows us to detect the target spin even if 
the inverse of $T_2^*$ is larger than 
the coupling from the target spin.
However, for the case when the reader is interested in how to eliminate 
the effect of the dipole-dipole interaction, we explain a scheme to 
nullify the dipole-dipole interaction between the probe 
spins for single spin detection with Dicke states in Appendix C.
In the rotating frame and under the rotating wave approximation, we can remove the terms oscillating with $\omega^{({\rm T})}$ and $\omega^{({\rm P})}$, and therefore the Hamiltonian in Eq.\ (\ref{eq:Hamiltonian}) effectively has only the Ising-type interaction
\begin{align}
\hat{H}^{({\rm eff})}=G\sum_{j=1}^L\frac{x_j^2+y_j^2-2z_j^2}{(x_j^2+y_j^2+z_j^2 )^{5/2}} \hat{\sigma}_z^{({\rm T})} \hat{\sigma}_{z,j}^{({\rm P})}.
\label{eq:effHamiltonian}
\end{align}
In this paper, we consider a case in which the target spin is either up or down, and so we replace $\hat{\sigma}_z^{({\rm T})}$ in Eq.\ (\ref{eq:effHamiltonian}) with a classical value $s=1$ or $-1$;
\begin{align}
\hat{H}^{({\rm eff})}_s&=\sum_{j=1}^L \frac{\omega_s(r_j,z_j)}{2} \hat{\sigma}_{z,j}^{({\rm P})},
\label{eq:effHamiltonians}\\
\omega_s(r_j,z_j)&=2Gs\times \frac{r_j^2-2z_j^2}{(r_j^2+z_j^2 )^{5/2}},
\end{align}
where we use the cylindrical coordinates $r=\sqrt{x^2+y^2}$ because of the rotational symmetry along the $z$ axis.
$\omega_s(r_j,z_j)$ denotes the inhomogeneous magnetic fields from the target spin, and  we show the $r$ and $z$ dependence of $\omega_s(r,z)/(2Gs)$ in Fig.\ 1 (c).
This graph shows that $\omega_s(r,z)/(2Gs)$ decreases as the distance from the origin increases.
We will estimate the parameter $s$ through the results of the readout using the probe spins with the Dicke state along the $x$ axis:
\begin{align}
\ket{D^L_{L/2}}_x=\tbinom{L}{L/2}^{-1/2}\sum_{{\rm perm}}(\ket{\underbrace{++\cdots +}_{L/2}\underbrace{--\cdots -}_{L/2}}),
\label{eq:squeezedstate}
\end{align}
where $\sum_{{\rm perm}}$ represents all permutations of the spins and $\ket{\pm}$ are the eigenstates of $ \sigma _x$.
For example, when $L=4$, we have $\ket{D^L_{L/2}}_x=\frac{1}{\sqrt{6}}(\ket{++--}+\ket{+-+-}+\ket{+--+}+\ket{-++-}+\ket{-+-+}+\ket{--++})$.
The probe spins with this state
are exposed to the inhomogeneous magnetic fields described
by the Hamiltonian Eq.\ (\ref{eq:effHamiltonians}).
Here, we assume the non-Markovian dephasing model, which is one of the most typical decoherences in solid-state systems \cite{matsuzaki2010quantum,yoshihara2006decoherence,kondo2016using,kakuyanagi2007dephasing,kalb2016experimental,hayashi2020experimental}.
The dynamics of the probe state under the effect of such a dephasing is given by the
following master equation:
\begin{align}
\frac{\partial \hat{\rho} (t)}{\partial t}=i[\hat{\rho} (t),\hat{H}^{({\rm eff})}_s]- \frac{t}{(T_2^*)^2}\sum_{j=1}^L\left(\hat{\rho} (t)-\hat{\sigma}_{z,j}^{(P)}\hat{\rho} (t)\hat{\sigma}_{z,j}^{(P)}\right),
\label{eq:Masterequation}
\end{align}
where $\hat{\rho} (t)$ is the density operator at time $t$ and $T_2^*$ denotes 
the time of free induction decay.
Throughout this paper, we do not consider the energy relaxation 
process characterized by
$T_1$. In the actual experiment, $T_1$ can be as 
long as $45$s at low temperature such as tens of millikelvin \cite{amsuss2011cavity,putz2017spectral}.
In our proposal, we use a superconducting flux qubit to create and measure Dicke states as we will describe below (see in section IV), and so we assume the use of a dilution refrigerator to keep the temperature around tens of millikelvin. In this assumption, the effect of  $T_1$ can be negligible.  
The first term of the right-hand side in Eq.\ (\ref{eq:Masterequation}) describes the interaction with the inhomogeneous magnetic fields from the target spin and the second term describes the decoherence. 

We describe the measurement sequence. 
First, prepare an initial state of the probe spins Eq.\ (\ref{eq:squeezedstate}). 
Second, let the quantum state evolve according to the master equation Eq.\ (\ref{eq:Masterequation}) for a time $t$. 
Third, measure the quantum state by a specific readout basis:
\begin{align}
\ket{{\rm Read}}=\frac{1}{\sqrt{2}}\left[\ket{D^L_{L/2}}_x+i\ket{D^L_{L/2+1}}_x\right],
\label{eq:Readout}
\end{align}
where $\ket{D^L_{L/2+1}}_x$ are also Dicke states defined as $\ket{D^L_{L/2+1}}_x=\tbinom{L}{L/2+1}^{-1/2}\sum_{{\rm perm}}(\ket{\underbrace{++\cdots +}_{L/2+1}\underbrace{--\cdots -}_{L/2-1}})$ ($\sum_{{\rm perm}}$ represents all permutations of the spins).
$\ket{{\rm Read}}$ represents the superposition of two Dicke states and is a kind of spin squeezed states \cite{ma2011quantum} (a similar state is analyzed in \cite{andre2002atom}).
Finally, repeat steps  1-3 $N$ times. 
We assume that the preparation time of the initial state and the readout time are negligibly small, and we can approximately obtain $N\simeq T/t$ where $T$ is a given total measurement time.

\section{Calculation of the sensitivity}
In this section, we show our results on the sensitivity to detect single spin with Dicke states.
We will explain the outline of the calculation of our results in the text and show the details of those derivations in Appendices \ref{appsec:explicit} and \ref{appsec:derivation}.

According to the prescription described above, we prepare an initial state $\hat{\rho}(0)=\ket{D^L_{L/2}}_x\bra{D^L_{L/2}}_x$, let this state evolve, and measure the state 
by the basis of $\ket{{\rm Read}}$, which provides us with a probability 
\begin{align}
p=\bra{{\rm Read}}\hat{\rho}(t)\ket{{\rm Read}}.
\label{eq:expectationvalue}
\end{align} 
The exact form of $p$ is described in Appendix \ref{appsec:explicit}. In order to estimate the uncertainty of the estimation of $s$ from the $N$ measurement values, we calculate the following 
\begin{align}
\delta s^{(\rm{Dicke})}:=\frac{\sqrt{p(1-p)}}{\sqrt{N}\left|\frac{\partial p}{\partial s}\right|},
\label{eq:Sensitivitydef}
\end{align}
where $\sqrt{p(1-p)}$ is the standard deviation of $p$.
Although 
the actual value of $s$ is discrete ($1$ or $-1$), we treat $s$ as a continuous variable
when we try to estimate it as follows. From the measurement $N$ results, we obtain an 
experimental value of $p$ with a finite variance, and we can estimate the value of $s$
from the information $p$ where we consider $s$ as a continuous variable. 
If the estimated value of $s$ is positive (negative), we expect that 
the actual value of $s$ is $+1$ ($-1$). 
In order to distinguish whether the target spin is up or down,
$\delta s^{(\rm{Dicke})}$ should be smaller than $1$. We will minimize $\delta s^{(\rm{Dicke})}$ by optimizing $t$ and the form of probe spins $z_{{\rm max}},r_{{\rm max}}$.
To rescale the time $t$, we set $t=uT_2^*/\sqrt{L}$ where $u$ denotes a dimensionless parameter. Throughout this paper, we only consider the limit of large $L$ and small $G$. In this assumption, we obtain
\begin{align}
\delta s^{(\rm{Dicke})}=\frac{F(u)}{\sqrt{TT_2^*}}\frac{L^{1/4}}{\left|\sum_j\frac{\partial\omega_s(r_j,z_j)}{\partial s}\right|}.
\label{eq:Sensitivity}
\end{align}
Here, the explicit form of $F(u)$ is shown in Eq.\ (\ref{eq:Fu})
and the derivation of $F(u)$ is shown in Appendix \ref{appsec:explicit}. 
Moreover, we calculate $\left|\sum_j\frac{\partial\omega_s(r_j,z_j)}{\partial s}\right|$ to take a continuous limit about the sum of the probe spins:
$\left|\sum_j\frac{\partial\omega_s(r_j,z_j)}{\partial s}\right|\simeq2G\rho\left|\iiint dxdydz\frac{r^2-2z^2}{(r^2+z^2 )^{5/2}}\right|
=4\pi G\rho\left|\frac{z_{{\rm max}}}{\sqrt{r^2+z_{{\rm max}}^2}}-\frac{z_{{\rm min}}}{\sqrt{r^2+z_{{\rm min}}^2}}\right|$.
This approximation is justified as $r_{{\rm max}},z_{{\rm max}},z_{{\rm min}}\gg \rho^{-1/3}$, where $\rho^{-1/3}$ is the average distance among each probe spin.
Finally, we optimize the form of the columnar substrate (that is, the number of the probe spins).
Using $L=\rho \pi r^2_{{\rm max}} (z_{{\rm max}}-z_{{\rm min}})$, we can obtain 
\begin{align}
\delta s_{{\rm min}}^{(\rm{Dicke})}=\frac{F(u_{{\rm min}})\times f_{{\rm min}}(\tilde{r}_{{\rm max}},\tilde{z}_{{\rm max}})}{4G\pi^{3/4}\sqrt{TT_2^*}}\frac{z_{{\rm min}}^{3/4}}{\rho^{3/4}},
\label{eq:Sensitivitymin}
\end{align} 
where $f(\tilde{r}_{{\rm max}},\tilde{z}_{{\rm max}})=[\tilde{r}^2_{{\rm max}} (\tilde{z}_{{\rm max}}-1)]^{1/4}\times \left(\frac{\tilde{z}_{{\rm max}}}{\sqrt{\tilde{r}^2_{{\rm max}}+\tilde{z}_{{\rm max}}^2}}-\frac{1}{\sqrt{\tilde{r}^2_{{\rm max}}+1}}\right)^{-1}$, and $\tilde{r}_{{\rm max}},\tilde{z}_{{\rm max}}$ are the normalized parameters $\tilde{r}_{{\rm max}}=r_{{\rm max}}/z_{{\rm min}}, \tilde{z}_{{\rm max}}=z_{{\rm max}}/z_{{\rm min}}$.
As a comparison, the explicit form of the single spin detection with separable states is given as follows \cite{uesugi2017single,hakoshima2020single}
\begin{eqnarray}
 \delta s^{({\rm sep})}_{{\rm min}}=\frac{\sqrt{2}e^{1/4}\times g_{{\rm min}}(\tilde{r}_{{\rm max}},\tilde{z}_{{\rm max}})}{4G\sqrt{\pi}\sqrt{TT_2^{*}}}\frac{z^{3/2}_{\rm min}}{\sqrt{\rho}},
\label{eq:Separablesensitivity}
\end{eqnarray}
where $g(\tilde{r}_{{\rm max}},\tilde{z}_{{\rm max}})=[\tilde{r}^2_{{\rm max}} (\tilde{z}_{{\rm max}}-1)]^{1/2}\times \left(\frac{\tilde{z}_{{\rm max}}}{\sqrt{\tilde{r}^2_{{\rm max}}+\tilde{z}_{{\rm max}}^2}}-\frac{1}{\sqrt{\tilde{r}^2_{{\rm max}}+1}}\right)^{-1}$.
According to \cite{uesugi2017single,hakoshima2020single}, this was numerically minimized as $g_{{\rm min}}(\tilde{r}_{{\rm max}},\tilde{z}_{{\rm max}})=5.32$ with $\tilde{r}_{{\rm max}}=0.928, \tilde{z}_{{\rm max}}=1.89$.
We can see that the scaling with $\rho $ and $z_{\rm{min}}$ for the entanglement scheme
is different from that of the separable scheme.
This is consistent with the previous results
where the entangled sensor to measure global homogeneous magnetic fields
 has a different scaling from that of
the separable states under
the effect of dephasing \cite{matsuzaki2011magnetic,chin2012quantum}.
In Table I, we show the summary of these scalings.
\begin{table}[htb]
\begin{tabular}{lccr}
\hline\hline Probe spins&$\delta s_{{\rm min}}$&$T_{\rm{s}}$&\begin{tabular}{c}
Preparation \\ \& \\ Readout\end{tabular}\\\hline
Separable \cite{uesugi2017single}&$O\left(\frac{z^{3/2}_{\rm min}}{\sqrt{\rho}}\right)$&$O\left(\frac{z^{3}_{\rm min}}{\rho T_2^{*}}\right)$&\begin{tabular}{r}Easy \\(but larger $\delta s_{{\rm min}}$)\end{tabular}\\
GHZ \cite{hakoshima2020single}&$O\left(\frac{z^{3/4}_{\rm min}}{\rho^{3/4}}\right)$&$O\left(\frac{z^{3/2}_{\rm min}}{\rho^{3/2} T_2^{*}}\right)$&\begin{tabular}{r}Difficult \\ (not scalable)\end{tabular}\\
\begin{tabular}{c}Dicke in Eq.~(\ref{eq:Sensitivitymin})\\ (our result)\end{tabular}&$O\left(\frac{z^{3/4}_{\rm min}}{\rho^{3/4}}\right)$&$O\left(\frac{z^{3/2}_{\rm min}}{\rho^{3/2} T_2^{*}}\right)$&\begin{tabular}{c}Section IV.\\ (scalable)\end{tabular}\\\hline\hline
\end{tabular}
\caption{Summary of the results. We compare three schemes of probe spins with separable states, GHZ states, and Dicke states. We show the scaling of $\delta s_{{\rm min}}$ and $T_{\rm{s}}$.  Also, we indicate whether the preparation and readout of these states are difficult or not for each scheme.}
\end{table}

In our expression of the uncertainty of the estimation of $s$, we need to minimize the functions of $f(\tilde{r}_{{\rm max}},\tilde{z}_{{\rm max}})$ and $F(u)$. 
Importantly, the form of $f(\tilde{r}_{{\rm max}},\tilde{z}_{{\rm max}})$ has been determined 
by the choice of the interaction time $t=uT_2^*/\sqrt{L}$ and the shape of the columnar substrate.
In the previous results on single spin detection with the GHZ states \cite{hakoshima2020single}, there was the same form 
as $f(\tilde{r}_{{\rm max}},\tilde{z}_{{\rm max}})$ in the sensitivity, and
this was numerically minimized as $f_{{\rm min}}(\tilde{r}_{{\rm max}},\tilde{z}_{{\rm max}})=4.14$ with $\tilde{r}_{{\rm max}}=1.87, \tilde{z}_{{\rm max}}=4.30$. 
We adopt the same minimization for our spin detection with Dicke states, and
we obtain the number of the probe spins as $L=\rho \pi \tilde{r}^2 (\tilde{z}_{{\rm max}}-1)z_{{\rm min}}^{3}=35.9\times \rho z_{{\rm min}}^{3}$.

On the other hand, we have derived $F(u)$ after we fix the initial state, the decoherence model, and the readout basis. 
Since $F(u)$ only depends on $u$, we can easily minimize it by a numerical method, and we obtain 
$F(u_{{\rm min}})=3.35$ when $u_{{\rm min}}=0.357$. It is worth mentioning that, if we replace $F(u)$ with $\sqrt{2}e^{1/4}=1.82$ in the expression $\delta s$, we obtain the uncertainty of $s$ when we use the GHZ state for the probe \cite{hakoshima2020single}. This means that, even if we use Dicke states that are 
experimentally feasible to realize, we can obtain a sensitivity comparable with the GHZ states that are typically considered as the best resource for quantum metrology.

To evaluate the performance of single spin detection with Dicke states, we will show the numerical result using realistic parameters.
We consider the nitrogen vacancy centers in diamond \cite{wolf2015subpicotesla,degen2008nanoscale,taylor2008high,muller2014nuclear,staudacher2013nuclear,maze2008nanoscale,balasubramanian2008nanoscale,lovchinsky2016nuclear,zhao2012sensing,shi2015single,abe2018tutorial,
schaffry2011proposed,degen2008scanning,mamin2013nanoscale,rugar2015proton,ohashi2013negatively,togan2010quantum,rondin2014magnetometry,grezes2015storage,balasubramanian2009ultralong,bauch2019decoherence} as probe spins.
According to the previous experiment \cite{bauch2019decoherence}, 
$T_2^{*}$ has a linear relation with $\rho^{-1}$, and the experimental value is
\begin{align}
\rho=&(1.98\times10^{12} {\rm~cm}^{-3}\cdot {\rm s})/T_2^{*} \\
&\qquad(10^{16} {\rm ~cm}^{-3}\lesssim \rho \lesssim 10^{19} {\rm~cm}^{-3})\notag.
\end{align}
By taking into account of the relation between $T_2^*$ and $\rho $,
we investigate how the sensitivity of the single spin detection changes by varying $\rho$. 
It is worth mentioning that, as the total measurement time $T$ increases, $\delta s$ decreases.
To quantify the performance of the single spin detection, we define the necessary measurement time $T=T_{\rm{s}}$ such that $\delta s=1$ should be satisfied.
If $T_{\rm{s}}$ is smaller, we can detect the target single spin for shorter measurement time, which is considered as a more efficient single spin detection scheme.
For the case of Dicke states of the probe spins, $T_{\rm{s}}$ is given by
\begin{align}
T_{\rm{s}}=&\frac{\left(F(u_{{\rm min}})\times f_{{\rm min}}(\tilde{r}_{{\rm max}},\tilde{z}_{{\rm max}})\right)^2}{16G^2\pi^{3/2}}\frac{z_{{\rm min}}^{3/2}}{T_2^*\rho^{3/2}}.
\label{eq:TsDicke}
\end{align}
\begin{figure}
\centering
  \includegraphics[clip,width=10cm]{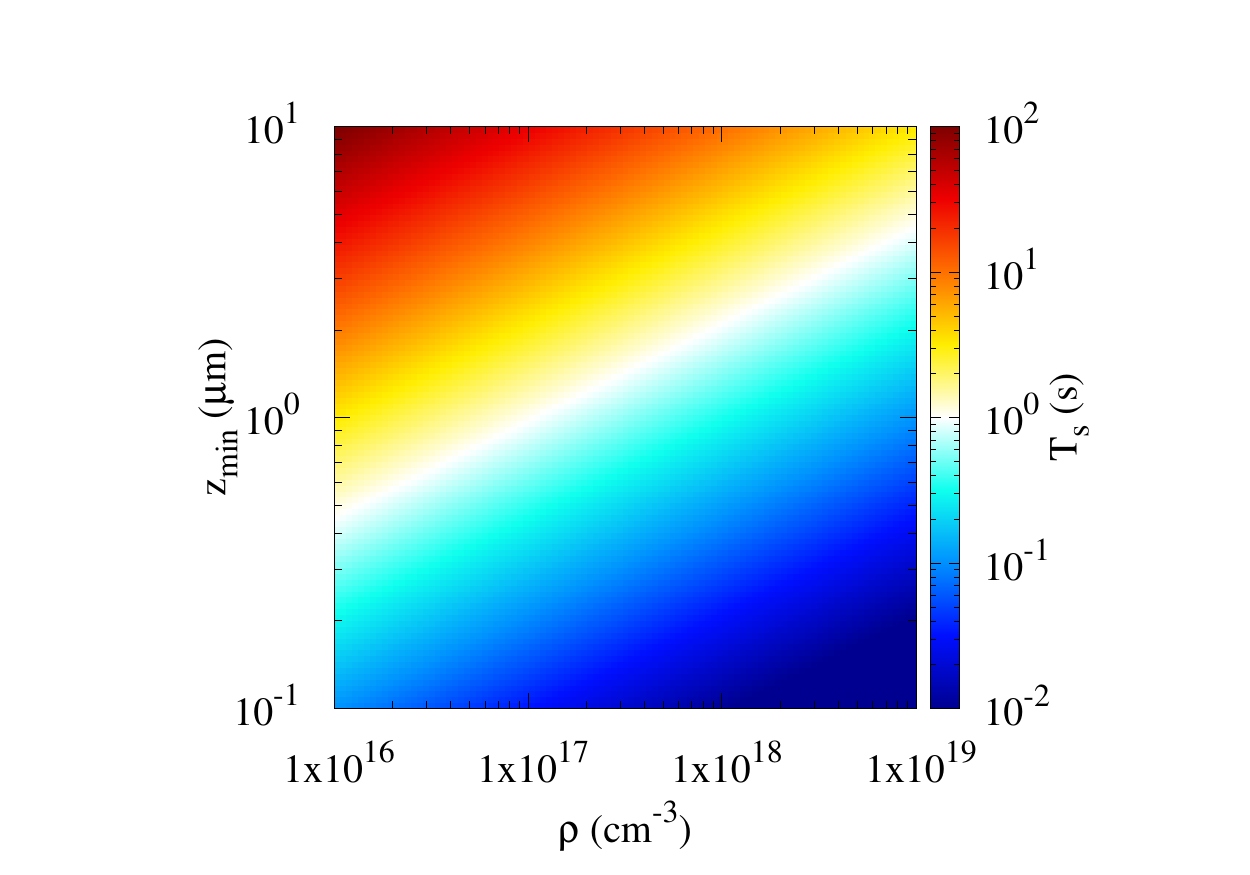}
  \caption{(color online) 
This graph shows the necessary measurement time $T_{\rm{s}}$ (s) against $\rho$ (cm$^{-3}$) and $z_{\rm{min}}$ for the case of Dicke states in Eq.~(\ref{eq:TsDicke}) of the probe spins.
We assume that the target spin is an electron spin.
}
\label{fig:Dickegraph1}
\includegraphics[clip,width=10cm]{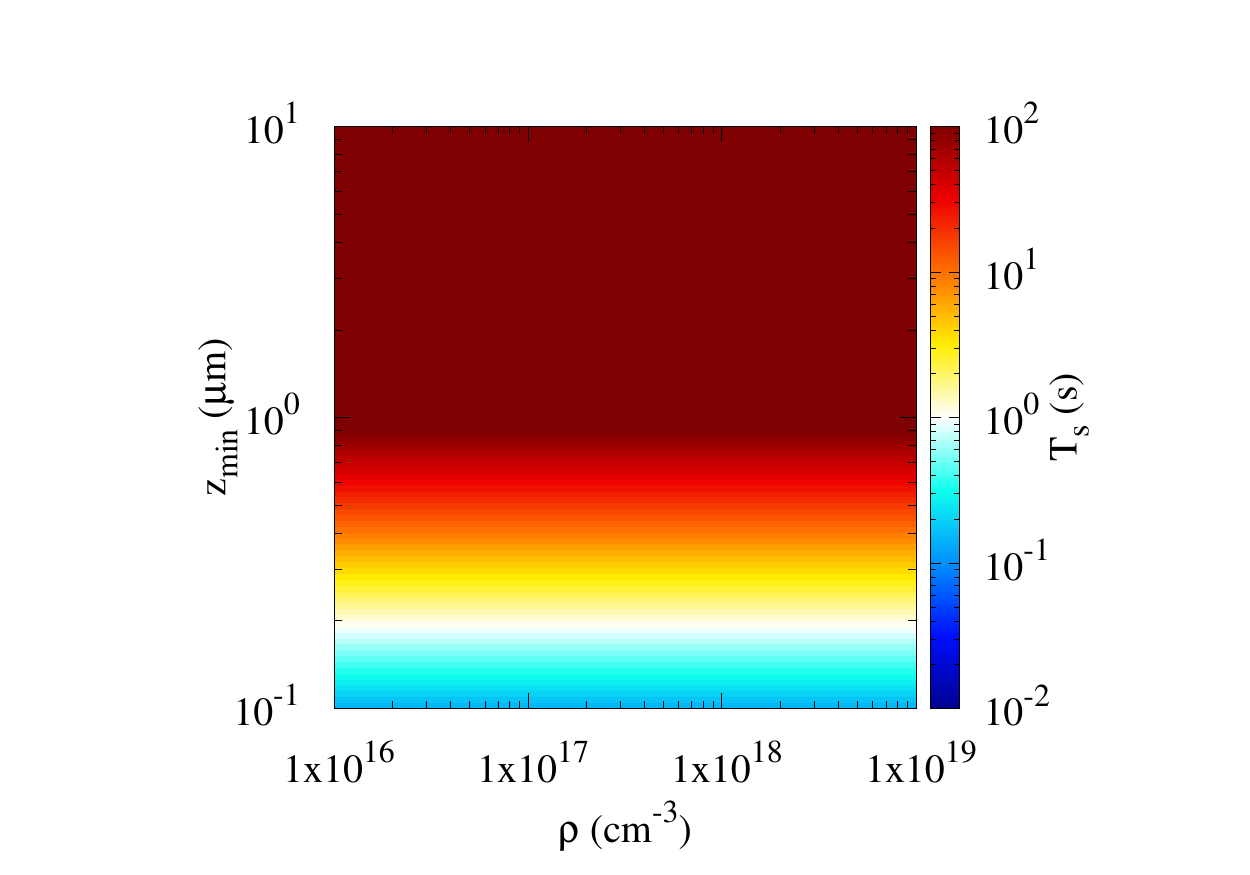}
  \caption{(color online) 
This graph shows the necessary measurement time $T_{\rm{s}}$ (s) against $\rho$ (cm$^{-3}$) and $z_{\rm{min}}$ for the case of the separable states of the probe spins.
}
\label{fig:Separablegraph1}
 \end{figure}
Fig.\ \ref{fig:Dickegraph1} and \ref{fig:Separablegraph1} show the detection time $T_{\rm{s}}$ against $\rho$ for the case of Dicke states and separable states \cite{uesugi2017single,hakoshima2020single}.
From this graph, $T_{\rm{s}}$ with Dicke states becomes smaller as $\rho$ increases with $z_{\rm{min}}$ fixed,
because $T_{\rm{s}}\propto \rho^{-1/2}z^{3/2}_{{\rm min}}$ from Eqs.\ (15) and (16).
On the other hand, $T_{\rm{s}}$ with separable states does not change because $\delta s^{({\rm sep})}$ in Eqs.\ (\ref{eq:Separablesensitivity}) and (15) depends only on $\rho T_2^{*}$ and therefore $T_{\rm{s}}\propto \rho^0z^{3}_{{\rm min}}$.

\section{Creation and measurement of Dicke states by a global and deterministic control}
In this subsection, 
we explain how to create and measure Dicke states by a global 
and deterministic control. For this purpose, we use another system 
that we call an ancillary qubit. In the actual setup, we could use a superconducting flux qubit for the ancillary qubit as we will describe below.
First, we consider the Hamiltonian of an ancillary qubit collectively coupled with many probe spins
\begin{align}
\hat{H}_{{\rm SS}}&=\hat{H}_{{\rm P}}+\hat{H}_{{\rm A}}+\hat{H}_{{\rm TR}}\\
\hat{H}_{{\rm P}}&=\omega^{({\rm P})}\hat{J}^{({\rm P})}_z, \quad \hat{H}_{{\rm A}}=\frac{\omega^{{\rm (A)}}}{2}\hat{\sigma}_z^{{\rm (A)}},\\
\hat{H}_{{\rm TR}}&=\lambda\sum_j^L \left(\hat{\sigma}_+^{{\rm (A)}}\hat{\sigma}_{-,j}^{(P)}+\hat{\sigma}_-^{{\rm (A)}}\hat{\sigma}_{+,j}^{(P)}\right),
\end{align}
where $\omega^{{\rm (A)}}$ and $\hat{\sigma}_z^{{\rm (A)}}$ are the resonant frequency 
and the Pauli $Z$ operator of the ancillary qubit, $\lambda$ 
denotes the transverse coupling strength between the ancillary qubit and the probe spins, and 
$\hat{J}^{({\rm P})}_z=\sum_{l=1}^{L}\hat{\sigma }_z^{({\rm P})}/2$.
$\hat{H}_{{\rm SS}}$ is the spin star model \cite{hutton2004mediated,breuer2004non}.
Moreover, we add the driving terms so as to perform the pulse operation 
\begin{align}
\hat{H}_{{\rm d}}&=\lambda_{{\rm d}}\hat{\sigma}_x^{{\rm (A)}}\cos{\omega^{({\rm d})}t}\\
\hat{H}_{{\rm d'}}&=\lambda_{{\rm d'}}\hat{J}^{({\rm P})}_x\cos{\omega^{({\rm d'})}t}
\end{align}
where $\omega^{({\rm d})}$ ($\omega^{({\rm d'})}$) denotes the frequency of driving fields for the ancillary qubit (probe spins), 
$\hat{J}^{({\rm P})}_x=\sum_{l=1}^{L}\hat{\sigma }_x^{({\rm P})}/2$ denotes the summation of the Pauli operators, 
and $\lambda_{{\rm d}}$ ($\lambda_{{\rm d'}}$) denotes the Rabi frequency for the ancillary qubit (probe spins). We assume $\omega^{({\rm A})}\gg \lambda ^{({\rm d})}$ and $\omega^{({\rm P})} \gg \lambda^{({\rm d'})}$. 
Also, we assume that we can turn on and off these Rabi frequencies. 
In our scheme, when we drive the ancillary spins (probe spin) by  setting a finite value of $\lambda_{{\rm d}}$($\lambda_{{\rm d'}}$), we turn of the driving off the probe spin (ancillary qubit) by setting $\lambda_{{\rm d'}}=0$ ($\lambda_{{\rm d}}$=0). 
We define that, if $\lambda_{{\rm d}}$ or $\lambda_{{\rm d'}}$ is much larger (smaller) than $\lambda$, we call it a hard (soft) pulse.
Intuitively, when we perform the hard pulses, the effect of the coupling between the ancillary qubit and probe spins is negligible during the pulse operations. 
It is known that this type of Hamiltonian was experimentally realized by a hybrid system composed of a superconducting qubit coupled with an electron-spin ensemble in diamond \cite{marcos2010coupling,kubo2011hybrid,zhu2011coherent,matsuzaki2015improving,matsuzaki2015improving}.
By using this Hamiltonian, we will show how to prepare the initial state of $\ket{D^L_{L/2}}_x$ and to readout the state with the basis of $\ket{{\rm Read}}$.

\subsubsection{Preparation of the initial state $\ket{D^L_{L/2}}_x$}
We show how to prepare the state $\ket{D^L_{L/2}}_x$.
The basic idea of our protocol is to repeat an energy transfer from the ancillary qubit to the probe spins with the flip-flop interaction.
Using Dicke states, $\hat{H}_{{\rm SS}}$ can be easily diagonalized.
Particularly when 
a resonant condition is satisfied
($\omega^{{\rm (A)}}=\omega^{{\rm (P)}}$), 
the energy eigenvalues and eigenstates are given by
\begin{align}
E_{n,\pm}&=\left(n-\frac{1}{2}\right)\omega^{{\rm (P)}}\pm\frac{1}{2}\mu_n\ \ \ (-L/2<n\leq L/2),\\
\mu_n&=2\lambda \sqrt{L/2(L/2+1)-n(n-1)},\\
\ket{E_{n,\pm}}&=\frac{1}{\sqrt{2}}\left(\ket{1}\ket{D^{L}_{n-1}}_z\pm\ket{0}\ket{D^{L}_{n}}_z\right),\\
E_{-L/2}&=-\frac{L+1}{2}\omega^{{\rm (P)}},\ \ \ E_{L/2+1}=\frac{L+1}{2}\omega^{{\rm (P)}},\\
\ket{E_{-L/2}} &= \ket{0}\ket{D^{L}_{0}}_z,\ \ \ \ket{E_{L/2+1}} = \ket{1}\ket{D^{L}_{L}}_z,
\end{align}
where $\ket{D^{L/2}_{n}}_z$ is the Dicke state along the $z$ axis $\ket{D^{L/2}_{n}}_z=\tbinom{L}{n}^{-1/2}\sum_{{\rm perm}}(\ket{\underbrace{\uparrow\uparrow\cdots \uparrow}_{n}\underbrace{\downarrow\downarrow\cdots \downarrow}_{L-n}})$.
Here, $n$ is the eigenvalue of $\hat{J}^{({\rm P})}_z$ for $-L/2<n\le L/2$.

\begin{figure}
\centering
  \includegraphics[clip,width=8.5cm]{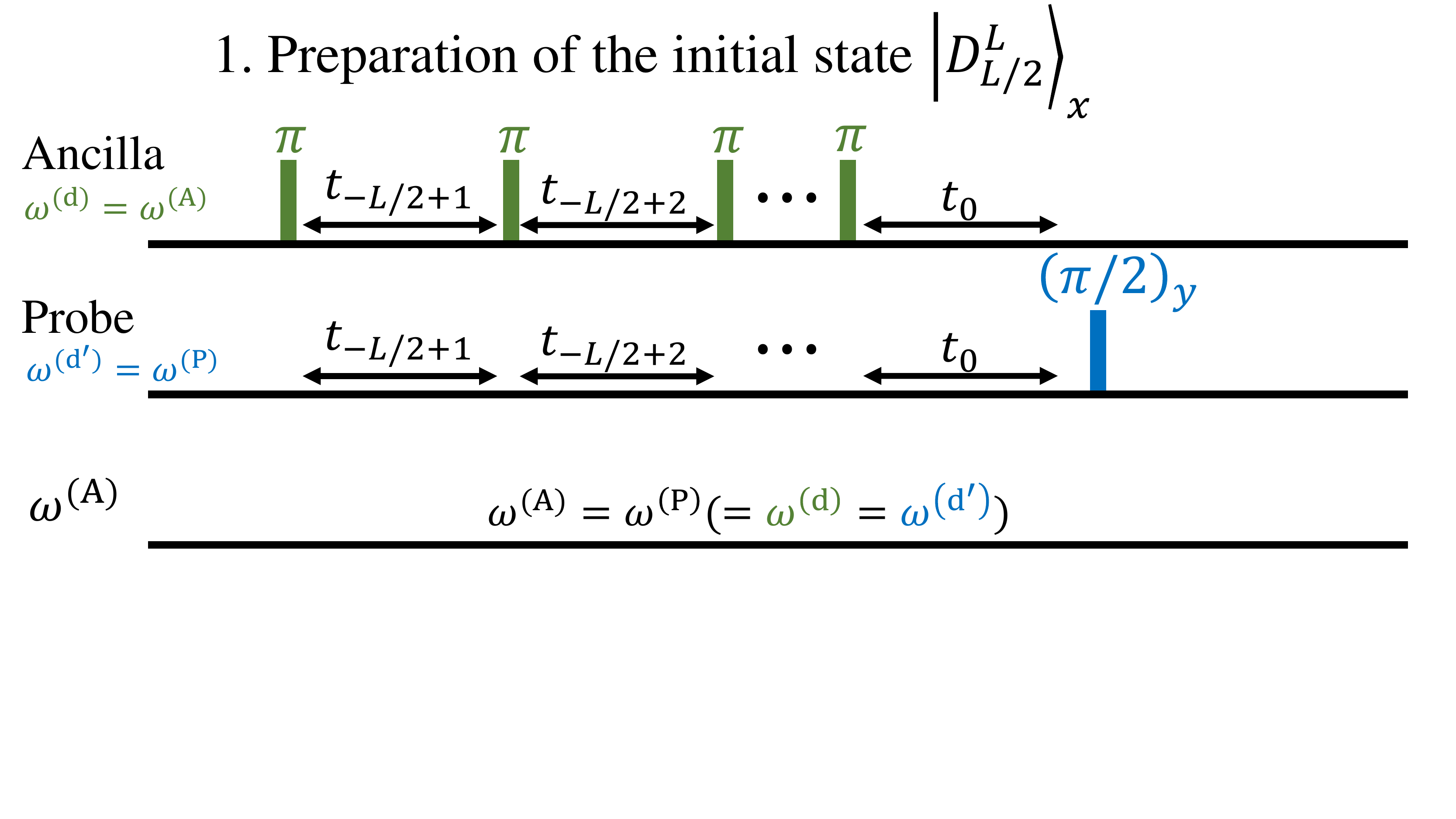}
  \caption{(color online) 
A schematic of the pulse sequence of the preparation of $\ket{D^{L}_{L/2}}_x$.
First, we perform a hard $\pi$ pulse to the ancillary qubit.
Second, let the system evolve by the Hamiltonian $\hat{H}_{{\rm SS}}$. Third, we repeat the first and second process 
$\frac{L}{2}$ times.
Finally, we perform a hard $\pi/2$ pulse along the $y$ axis into the probe spin ensemble.}
\label{fig:PulseSequence1}
 \end{figure}
Fig.\ \ref{fig:PulseSequence1} shows the
 pulse sequence of the preparation of $\ket{D^{L}_{L/2}}_x$.
In step 1, we prepare an initial state
\begin{align}
\ket{E_{-L/2}}  =\ket{0}\ket{D^{L}_{0}}_z.
\end{align}
In step 2, we excite the ancillary qubit by a hard $\pi $ pulse
\begin{align}
\ket{1}\ket{D^{L}_{0}}_z=e^{-i\pi\sigma_y^{{\rm (A)}}/2}\ket{0}\ket{D^{L}_{0}}_z,
\end{align}
which can be realized by turning on $\lambda^{({\rm d})}$ and choosing $\omega^{{\rm (d)}}=\omega^{{\rm (A)}}$.
In step 3, let the system evolve by the Hamiltonian $\hat{H}_{{\rm SS}}$ for a certain time until the excitation of the ancillary qubit is completely transferred to the spin ensemble, and we obtain 
\begin{align}
\ket{0}\ket{D^{L/2}_{1}}_z=\exp{[-i\hat{H}'t_{-L/2+1}]}\ket{1}\ket{D^{L}_{0}}_z.
\end{align}
[The interaction time is $t_{-L/2+1}=\pi/(E_{-L/2+1,+}-E_{-L/2+1,-})=\pi/\mu_{-L/2+1,-}$ in this case.] 
In step 4, repeat steps 2 and 3 by changing the evolution time $t_{-m}$ for the energy excitation transfer: 
\begin{align}
\ket{0}\ket{D^{L}_{L/2}}_z=\prod_{m=0}^{L/2-1}\left(\exp{[-i\hat{H}'t_{-m}]}e^{-i\pi\sigma_y^{{\rm (A)}}/2}\right)\ket{0}\ket{D^{L}_{0}}_z,
\label{eq:Dickezcreation}
\end{align}
where $t_{-m}=\pi/(E_{-m+1,+}-E_{-m+1,-})=\pi/\mu_{-m+1}$. We repeat these steps $L/2$ times.
In step 5, by turning on $\lambda_{{\rm d'}}$ in order to perform a hard $\pi/2$ pulse along the $y$ axis into the probe spin ensemble, we obtain
\begin{align}
\ket{0}\ket{D^{L}_{L/2}}_x=e^{-i\pi\hat{J}_y^{{\rm (P)}}/2}\ket{0}\ket{D^{L}_{L/2}}_z.
\end{align}

\subsubsection{Readout by $\ket{{\rm Read}}$}
We show how to readout the state with the basis of $\ket{{\rm Read}}$.
If we can construct a unitary operator $U_{{\rm Read}}$ as $\ket{{\rm Read}}=U_{{\rm Read}}\ket{D^{L}_{0}}_z$, the expectation value $p$ [already defined by Eq. (\ref{eq:expectationvalue})] can be rewritten as
\begin{align}
p=\bra{D^{L}_{0}}_zU_{{\rm Read}}^\dagger\hat{\rho}(t)U_{{\rm Read}}\ket{D^{L}_{0}}_z.
\end{align}
This means that the combination of the inverse operation $U_{{\rm Read}}^\dagger$ and the global projection measurement to all-down state $\ket{D^{L}_{0}}_z=\ket{\downarrow\cdots\downarrow}$ provides us with a way to obtain
the value of $p$, which is the probability to measure the state with the basis of $\ket{{\rm Read}}$. So we consider how to construct $U_{{\rm Read}}^\dagger$. 

The construction of $U_{{\rm Read}}$ is as follows.
The basic idea is to use $\ket{0}\ket{D^{L}_{L/2}}_z$ and $\ket{0}\ket{D^{L}_{L/2+1}}_z$
as an effective qubit due to the frequency selectivity.
If a resonant condition is not satisfied, 
$\omega^{{\rm (A)}}\gg\omega^{{\rm (P)}}$ and $\lambda$, we can obtain the 
effective Hamiltonian as follows:
\begin{align}
\hat{H}_{{\rm SS}}^{{\rm (eff)}}=\hat{H}_{{\rm P}}+\hat{H}_{{\rm A}}-\frac{\lambda^2}{\omega^{{\rm (A)}}-\omega^{{\rm (P)}}}\hat{\sigma}_z^{{\rm (A)}}  (\hat{J}_z^{{\rm (P)}})^2.
\end{align}
Here, 
the energy eigenstates of $\hat{H}_{{\rm SS}}^{{\rm (eff)}}$ are expressed by the separable states of $\ket{0}$ ($\ket{1}$) and Dicke states such as $\ket{0}\ket{D^{L}_{L/2+1}}_z$ or $\ket{0}\ket{D^{L}_{L/2-1}}_z$.
The difference of the eigenvalues between $\ket{0}\ket{D^{L}_{L/2}}_z$ and $\ket{0}\ket{D^{L}_{L/2+1}}_z$ 
is $\omega^{{\rm (P)}}+\frac{\lambda^2}{\omega^{{\rm (A)}}-\omega^{{\rm (P)}}}$, which is detuned from other 
energy eigenstates.
 For example, 
the difference of the eigenvalues between $\ket{0}\ket{D^{L}_{L/2}}_z$ and $\ket{0}\ket{D^{L}_{L/2-1}}_z$ is $\omega^{{\rm (P)}}-\frac{\lambda^2}{\omega^{{\rm (A)}}-\omega^{{\rm (P)}}}$.

\begin{figure}
\centering
  \includegraphics[clip,width=8.5cm]{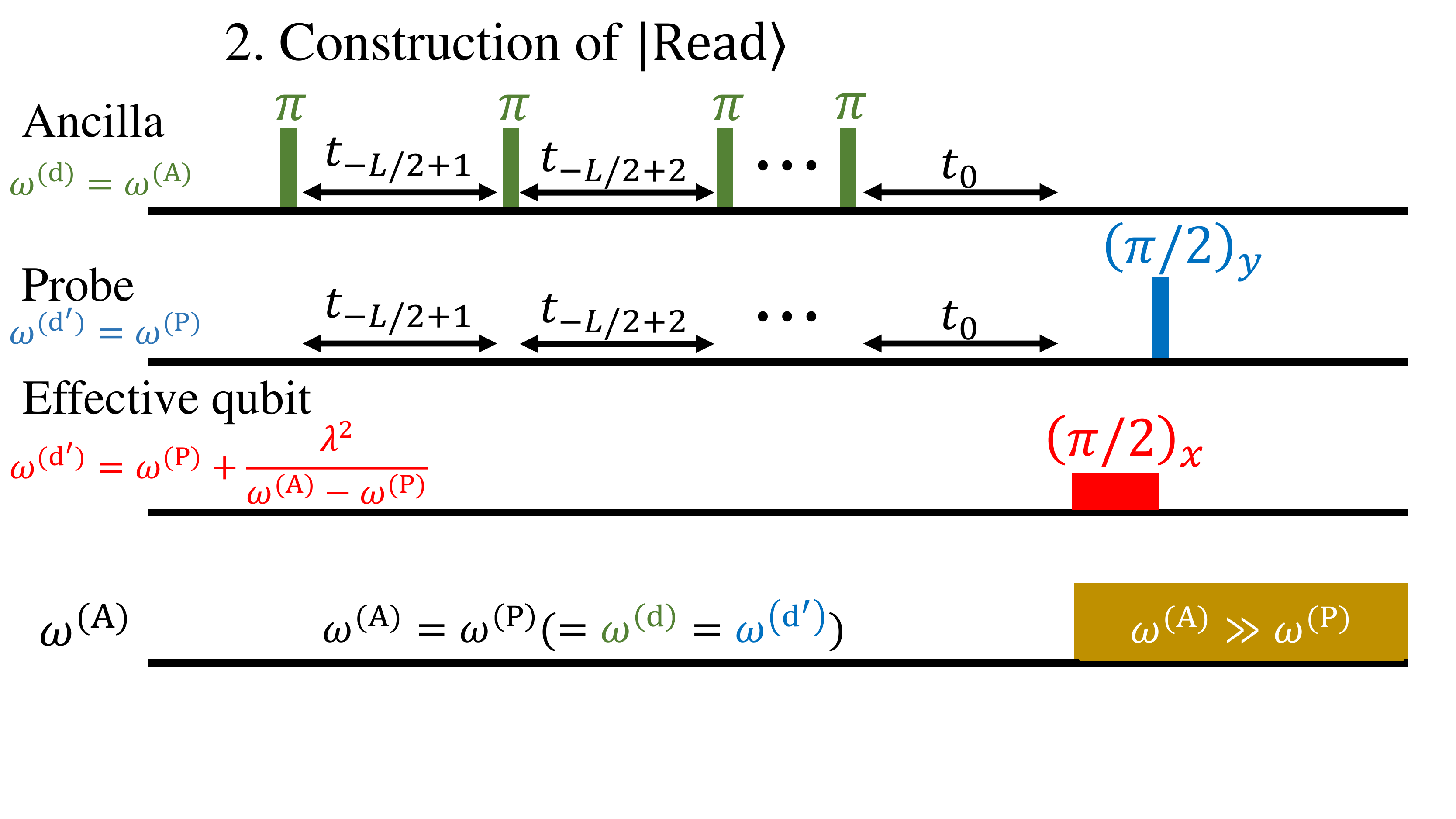}
  \caption{(color online) 
A schematic of
the pulse sequence of the construction of $\ket{{\rm Read}}$.
First,
we perform a hard $\pi$ pulse to the ancillary qubit. 
Second, let the system evolve by the Hamiltonian $\hat{H}_{{\rm SS}}$. Thirdly, repeat the second and third process 
$\frac{L}{2}$ times.
Fourth, we increase the frequency of the ancillary qubit to induce the detuning
and globally perform the soft $\pi/2$ pulse to the spin ensemble with $\omega^{{\rm (d')}}=\omega^{{\rm (P)}}+\frac{\lambda^2}{\omega^{{\rm (A)}}-\omega^{{\rm (P)}}}$.
Finally, we perform a hard $\pi/2$ pulse along the $y$ axis into the probe spin ensemble. 
}
\label{fig:PulseSequence2}
 \end{figure}
Fig.\ \ref{fig:PulseSequence2} shows the pulse sequence of the construction of $\ket{{\rm Read}}$.
First, we prepare the state in Eq.\ (\ref{eq:Dickezcreation}).
Second, we globally perform the soft $\pi/2$ pulse to the spin ensemble by turning on $\lambda_{{\rm d'}}$ with $\omega^{{\rm (d')}}=\omega^{{\rm (P)}}+\frac{\lambda^2}{\omega^{{\rm (A)}}-\omega^{{\rm (P)}}}$
\begin{align}
\frac{1}{\sqrt{2}}\ket{0}\left(\ket{D^{L}_{L/2}}_z+i\ket{D^{L}_{L/2+1}}_z\right)=U_{{\rm pulse}}\ket{0}\ket{D^{L}_{L/2}}_z.
\end{align}
Finally, by turning off $\lambda$ and choosing $\omega^{{\rm (d')}}=\omega^{{\rm (P)}}$ in order to perform the hard $\pi/2$ pulse along the $y$ axis into each of the probe spins, we obtain 
\begin{align}
\ket{0}\ket{{\rm Read}}=\frac{e^{-i\pi\hat{J}_y^{{\rm (P)}}/2}}{\sqrt{2}}\ket{0}\left(\ket{D^{L}_{L/2}}_z+i\ket{D^{L}_{L/2+1}}_z\right),
\end{align}
and  
\begin{align}
U_{{\rm Read}}=e^{-i\pi\hat{J}_y^{{\rm (P)}}/2}U_{{\rm pulse}}\prod_{m=0}^{L/2-1}\left(\exp{[-i\hat{H}'t_{-m}]}e^{-i\pi\sigma_y^{{\rm (A)}}/2}\right).
\end{align}

\section{Summary and Conclusion}
To summarize, we propose single spin detection by using Dicke states as probes, and evaluate its performance.
Particularly, we investigate the necessary time $T_{\rm{s}}$ to readout the target spin
with a probe of Dicke states
and we compare it with that of the classical strategy where only separable states are used as the probe.
Assuming a relationship of $\rho\propto (T_2^*)^{-1}$ 
(which has been experimentally observed in some systems), we show that 
$T_{\rm{s}}$ becomes smaller as $\rho$ increases for the case of Dicke states, while $T_{\rm{s}}$ does not depend on $\rho$
for the classical strategy. Therefore, we conclude that 
by using dense probe spins, Dicke states provide higher sensitivity than separable states when we aim to detect 
a single spin.
Moreover, we propose how to create and measure
 Dicke states by a global and deterministic control. Our results pave the way for a rapid spin detection that is useful
for many areas such as condensed-matter physics, material science, and life sciences.

\begin{acknowledgments}
We are grateful to Junko Ishi-Hayase for assistance in this study. 
This work was supported by the Leading Initiative for Excellent Young Researchers of the Ministry of Education, Culture,
Sports, Science, and Technology (MEXT), Japan, under KAKENHI Grant No. 15H05870; and by the  JSTJapan Science and Technology Agency, Japan,
under Presto Grant No. JPMJPR1919.
\end{acknowledgments}
\begin{widetext}
\appendix

\section{The explicit form of $p$ and $F(u)$}\label{appsec:explicit}
Using the explicit form of $p$ (see the derivation in Appendix \ref{appsec:derivation}), we can derivate $F(u)$ in Eq.\ (\ref{eq:Sensitivity}).
The explicit form of $p$ is given by
\begin{align}
p&=\frac{e^{-\frac{u^2}{2}}}{2}I_{0}(u^2/4)\left[I_{0}(u^2/4)+I_{1}(u^2/4)\right]+\frac{ T_2}{2\sqrt{L}}ue^{-\frac{u^2}{2}}\frac{1}{2}\left(I_{0}(u^2/4)-I_{1}(u^2/4)\right)^2\left(\sum_j\omega_s(r_j,z_j)\right),
\label{appeq:explicitformp}
\end{align}
where $I_{\alpha}(x)$ is the modified Bessel function.
From this, we obtain
\begin{align}
p(1-p)&\simeq\frac{e^{-\frac{u^2}{2}}}{2}I_{0}(u^2/4)\left[I_{0}(u^2/4)+I_{1}(u^2/4)\right]\left(1-\frac{e^{-\frac{u^2}{2}}}{2}I_{0}(u^2/4)\left[I_{0}(u^2/4)+I_{1}(u^2/4)\right]\right),\\
\left|\frac{\partial p}{\partial s}\right|&=\frac{ T_2}{2\sqrt{L}}ue^{-\frac{u^2}{2}}\frac{1}{2}\left(I_{0}(u^2/4)-I_{1}(u^2/4)\right)^2\left|\sum_j\frac{\partial \omega_s(r_j,z_j)}{\partial s}\right|,
\end{align}
and $\sqrt{N}=\sqrt{T/(T_2u/\sqrt{L})}$; we can derive Eq.\ (\ref{eq:Sensitivity}) by using
\begin{align}
F(u)=\frac{2\sqrt{2I_{0}(\frac{u^2}{4})\left(I_{0}(\frac{u^2}{4})+I_{1}(\frac{u^2}{4})\right)\left(1-\frac{e^{-\frac{u^2}{2}}I_{0}(\frac{u^2}{4})\left(I_{0}(\frac{u^2}{4})+I_{1}(\frac{u^2}{4})\right)}{2}\right)}}{\sqrt{u}e^{-\frac{u^2}{4}}\left(I_{0}(\frac{u^2}{4})-I_{1}(\frac{u^2}{4})\right)^2}.
\label{eq:Fu}
\end{align}
\section{Derivation of the expectation value $p$ in Eq.\ (\ref{appeq:explicitformp})}\label{appsec:derivation}
From Eq.\ (\ref{eq:expectationvalue}), $p$ can be rewritten as
\begin{align}
p&=\frac{1}{2}\bra{D^{L}_{L/2}}_x\hat{\rho} (t)\ket{D^{L}_{L/2}}_x+\frac{1}{2}\bra{D^{L}_{L/2+1}}_x\hat{\rho} (t)\ket{D^{L}_{L/2+1}}_x-\mathrm{Im}\left[\bra{D^{L}_{L/2}}_x\hat{\rho} (t)\ket{D^{L}_{L/2+1}}_x\right],
\end{align}
where $\mathrm{Im}[\cdot]$ denotes the imaginary part.
We will calculate these three terms.

Here, we rewrite $\ket{D^{L}_{L/2}}_x$ with the basis of $\ket{0}$ or $\ket{1}$:
\begin{align}
\ket{D^{L}_{L/2}}_x&=\frac{1}{2^{\frac L 2}}\sum_m \zeta(m)\ket{m},\\
\zeta(m)&=\frac{1}{\sqrt{\tbinom{L}{L/2}}}\sum_{j}(-1)^{\braket{j,m}},\\
\braket{j,m}&=j_1m_1+j_2m_2+\cdots +j_Lm_L,\\
j&=j_1j_2\cdots j_L= \underbrace{00\cdots0}_{L/2}\underbrace{11\cdots1}_{L/2},\qquad m=m_1m_2\cdots m_L,
\end{align}
where $m=m_1m_2\cdots m_L$ is a $L$ bit sequence and $m_i=0$ or $1$ ($i=1,\cdots, L$) denotes the eigenvalues of $\sigma_{z,i}$, and $j=j_1j_2\cdots j_L$ ($j_i=0$ or $1$, $i=1,\cdots, L$).
We assume that half of the components of $j$ are $1$ and the other half of the components of $j$ are zero.
$\sum_m$ denotes $\sum_{m_1,m_2,\cdots, m_L}$ (the sum of the $2^L$ terms), and $\sum_{j}$ denotes all the permutations of $j$ corresponding to Eq.\ (\ref{eq:squeezedstate}) [the sum of the $\tbinom{L}{L/2}$ terms], which means all permutations such that half of the components of $j$ are $1$ and the other half of the components $j$ are zero.
For example, when $L=4$, all the permutation of $j$ are $j=0011,0101,0110,1001,1010,1100$, and when $j=0011$, $j_1=j_2=0$, and $j_3=j_4=1$.
In this case, $\ket{D^{L}_{L/2}}_x=\frac{1}{2\sqrt{6}}\left(3(\ket{0000}+\ket{1111})-(\ket{0011}+\ket{0101}+\ket{0110}+\ket{1001}+\ket{1010}+\ket{1100})\right)$.
Moreover, we rewrite $\ket{D^{L}_{L/2+1}}_x$ with the basis of $\ket{0}$ or $\ket{1}$:
\begin{align}
\ket{D^{L}_{L/2+1}}_x&=\frac{1}{2^{\frac L 2}}\sum_m \xi(m)\ket{m},\\
\xi(m)&=\frac{1}{\sqrt{\tbinom{L}{L/2+1}}}\sum_{l}(-1)^{\braket{l,m}},\\
l&= \underbrace{00\cdots0}_{L/2-1}\underbrace{11\cdots1}_{L/2+1},
\end{align}
where $l=l_1l_2\cdots l_L$ and $\sum_{l}$ denotes all the permutations of $l$ [the sum of the $\tbinom{L}{L/2+1}$ terms].
For example, when $L=4$, all the permutation of $l$ are $l=0111,1011,1101,1110$, and when $l=0111$, $l_1=0$, and $l_2=l_3=l_4=1$.
In this case, $\ket{D^{L}_{L/2+1}}_x=\frac{1}{4}\left(2(\ket{0000}-\ket{1111})+(\ket{0001}+\ket{0010}+\ket{0100}+\ket{1000}+\ket{0111}+\ket{1011}+\ket{1101}+\ket{1110})\right)$.

The solution of Eq.\ (\ref{eq:Masterequation}) is given by
\begin{align}
\hat{\rho} (t)&=\frac{1}{2^{L}}\sum_{m,m'} \zeta(m)\zeta(m')\ket{m}\bra{m'}\times \exp{\left[i\sum_n \frac{\omega_s(r_n,z_n) }{2}t\{(-1)^{m_n}-(-1)^{m_n'}\}\right]}\times\prod_n\left(\delta_{m_n,m_n'}+(1-\delta_{m_n,m_n'})e^{-\left(\frac{t}{T_2}\right)^2}\right),
\end{align}
where the $\exp{[\cdots]}$ term expresses the unitary time evolution and the $\prod_n(\cdots)$ term expresses the decoherence corresponding to the first and second term in Eq.\ (\ref{eq:Masterequation}), respectively.

\subsection{First term calculation: $\bra{D^{L}_{L/2}}_x\hat{\rho} (t)\ket{D^{L}_{L/2}}_x$}
The first term $\bra{D^{L}_{L/2}}_x\hat{\rho} (t)\ket{D^{L}_{L/2}}_x$ gives
\begin{align}
&\frac{1}{2^{2L}}\sum_{m,m'} \zeta(m)^2\zeta(m')^2\times \exp{\left[i\sum_n \frac{\omega_s(r_n,z_n) }{2}t\{(-1)^{m_n}-(-1)^{m_n'}\}\right]}\times\prod_n\left(\delta_{m_n,m_n'}+(1-\delta_{m_n,m_n'})e^{-\left(\frac{t}{T_2}\right)^2}\right)\\
&=\frac{1}{2^{2L}\tbinom{L}{L/2}^2}\sum_{m,m'}\sum_{j^{(1)}j^{(2)}j^{(3)}j^{(4)}}(-1)^{\braket{j^{(1)}+j^{(2)},m}+\braket{j^{(3)}+j^{(4)},m'}} e^{\left[i\sum_n \frac{\omega_s(r_n,z_n) }{2}t\{(-1)^{m_n}-(-1)^{m_n'}\}\right]}\prod_n\left(\delta_{m_n,m_n'}+(1-\delta_{m_n,m_n'})e^{-\left(\frac{t}{T_2}\right)^2}\right)\\
&=\frac{1}{2^{2L}\tbinom{L}{L/2}^2}\sum_{j^{(1)}j^{(2)}j^{(3)}j^{(4)}}\prod_n\left(\sum_{m_n,m_n'=0}^1(-1)^{(j^{(1)}_n+j^{(2)}_n)m_n+(j^{(3)}_n+j^{(4)}_n)m_n'}e^{\left[i\sum_n \omega_s(r_n,z_n)t\frac{(-1)^{m_n}-(-1)^{m_n'}}{2}\right]}\left(\delta_{m_n,m_n'}+(1-\delta_{m_n,m_n'})e^{-\left(\frac{t}{T_2}\right)^2}\right)\right)\\
&=\frac{1}{2^{2L}\tbinom{L}{L/2}^2}\sum_{j^{(1)}j^{(2)}j^{(3)}j^{(4)}}\prod_n\left(1+(-1)^{j^{(1)}_n+j^{(2)}_n+j^{(3)}_n+j^{(4)}_n}+(-1)^{j^{(1)}_n+j^{(2)}_n}e^{i\omega_s(r_n,z_n) t}e^{-\left(\frac{t}{T_2}\right)^2}+(-1)^{j^{(3)}_n+j^{(4)}_n}e^{-i\omega_s(r_n,z_n) t }e^{-\left(\frac{t}{T_2}\right)^2}\right)\\
&=\frac{1}{2^{L}\tbinom{L}{L/2}^2}\sum_{j^{(1)}j^{(2)}j^{(3)}j^{(4)}}\prod_n\left(\delta_{j^{(1)}_n+j^{(2)}_n+j^{(3)}_n+j^{(4)}_n \equiv 0}+\delta_{j^{(1)}_n,j^{(2)}_n}e^{i\omega_s(r_n,z_n) t }e^{-\left(\frac{t}{T_2}\right)^2}+\delta_{j^{(3)}_n,j^{(4)}_n}e^{-i\omega_s(r_n,z_n) t }e^{-\left(\frac{t}{T_2}\right)^2}-\cos{\omega_s(r_n,z_n) t }e^{-\left(\frac{t}{T_2}\right)^2}\right)\\
&=\frac{1}{2^{L}\tbinom{L}{L/2}^2}\sum_{j^{(1)}j^{(2)}j^{(3)}j^{(4)}}\prod_n\left(\delta_{j^{(1)}_n+j^{(2)}_n+j^{(3)}_n+j^{(4)}_n \equiv 0}+\delta_{j^{(1)}_n,j^{(2)}_n}(1+i\omega_s(r_n,z_n) t )e^{-\left(\frac{t}{T_2}\right)^2}+\delta_{j^{(3)}_n,j^{(4)}_n}(1-i\omega_s(r_n,z_n) t )e^{-\left(\frac{t}{T_2}\right)^2}-e^{-\left(\frac{t}{T_2}\right)^2}\right)\notag\\
&\qquad\qquad\qquad\qquad\qquad\qquad\qquad\qquad\qquad\qquad\qquad\qquad\qquad\qquad\qquad\qquad\qquad\qquad\qquad+O((\omega_s(r_n,z_n) t)^2).
\label{appeq:firstterm}
\end{align}
Here, $\delta_{j^{(1)}_n+j^{(2)}_n+j^{(3)}_n+j^{(4)}_n \equiv 0}=1$ (or $0$) if $j^{(1)}_n+j^{(2)}_n+j^{(3)}_n+j^{(4)}_n \equiv 0$ (or $1$) $\pmod{2}$.
We assume that $O((\omega_s(r_n,z_n) t)^2)$ is negligibly small.
Table \ref{table:fourcasesoffirstterm} shows four cases of the contents of $\prod_n[\cdots]$ in Eq.\ (\ref{appeq:firstterm}).
\begin{table}[htb]
\centering
\caption{Four cases of the contents of $\prod_n[\cdots]$ in Eq.\ (\ref{appeq:firstterm})}
  \begin{tabular}{|c|c|} \hline
Four cases&values \\\hline\hline
    $j^{(1)}_n+j^{(2)}_n\equiv 0$, $j^{(3)}_n+j^{(4)}_n\equiv 0$ $\left(j^{(1)}_n+j^{(2)}_n+j^{(3)}_n+j^{(4)}_n\equiv 0\right)$&$1+e^{-\left(\frac{t}{T_2}\right)^2}$ \\ \hline 
    $j^{(1)}_n+j^{(2)}_n\equiv 0$, $j^{(3)}_n+j^{(4)}_n\equiv 1$ $\left(j^{(1)}_n+j^{(2)}_n+j^{(3)}_n+j^{(4)}_n\equiv 1\right)$&$(i\omega_s(r_n,z_n) t) e^{-\left(\frac{t}{T_2}\right)^2}$ \\ \hline 
    $j^{(1)}_n+j^{(2)}_n\equiv 1$, $j^{(3)}_n+j^{(4)}_n\equiv 0$ $\left(j^{(1)}_n+j^{(2)}_n+j^{(3)}_n+j^{(4)}_n\equiv 1\right)$&$(-i\omega_s(r_n,z_n) t )e^{-\left(\frac{t}{T_2}\right)^2}$ \\ \hline 
    $j^{(1)}_n+j^{(2)}_n\equiv 1$, $j^{(3)}_n+j^{(4)}_n\equiv 1$ $\left(j^{(1)}_n+j^{(2)}_n+j^{(3)}_n+j^{(4)}_n\equiv 0\right)$&$1-e^{-\left(\frac{t}{T_2}\right)^2}$ \\ \hline 
  \end{tabular}
\label{table:fourcasesoffirstterm}
\end{table}
From Eq.\ (\ref{appeq:firstterm}), we have a term of $\delta_{j^{(1)}_n,j^{(2)}_n}(i\omega_s(r_n,z_n) t )$ and also a term of $\delta_{j^{(3)}_n,j^{(4)}_n}(-i\omega_s(r_n,z_n) t )$. 
After the summation of $j^{(1)},j^{(2)},j^{(3)},j^{(4)}$, these terms cancel each other so that we should not have a term of $O(\omega_s(r_n,z_n) t)$. 
Therefore, in Table \ref{table:fourcasesoffirstterm}, we can just consider the first line and fourth line. 
This means that we can consider only the following condition: 
\begin{align}
j^{(1)}_n+j^{(2)}_n+j^{(3)}_n+j^{(4)}_n\equiv 0 \pmod{2}\qquad ({\rm for\, all\, }n).
\label{appeq:followingcondition}
\end{align}
We need to count how many sets of $j^{(1)},j^{(2)},j^{(3)}$, and $j^{(4)}$ exist to satisfy the condition of Eq.\ (\ref{appeq:followingcondition}).

First, we fix the sequence $j^{(1)}$ to
\begin{align}
j^{(1)}=\underbrace{000\cdots 0}_{L/2}\underbrace{111\cdots 1}_{L/2}
\end{align}
Second, we consider the sequences $j^{(2)}$ which satisfy the condition that the sequences of $j^{(1)}+j^{(2)}$ contain a $L-2n$ number of zero and a $2n$ number of $1$.
For example, when
\begin{align}
j^{(2)}=\underbrace{000\cdots 0}_{L/2-n}\underbrace{111\cdots 1}_{n}\underbrace{000\cdots 0}_{n}\underbrace{111\cdots 1}_{L/2-n},
\end{align}
we obtain
\begin{align}
j^{(1)}+j^{(2)}\equiv\underbrace{000\cdots 0}_{L/2-n}\underbrace{111\cdots 1}_{n}\underbrace{111\cdots 1}_{n}\underbrace{000\cdots 0}_{L/2-n},
\end{align}
and this sequence surely contains a $L-2n$ number of zero and a $2n$ number of $1$.
Since $j^{(1)}$ is fixed, let us consider how many configurations of $j^{(2)}$ are possible. 
Of course, the total number of configurations of $j^{(2)}$ is $\tbinom{L}{L/2}$. 
However, we consider a condition such that the number of $1$ should be $n$ in the left side, as seen in Eq.\ (B18). 
In this condition, the number of possible configurations of $j^{(2)}$ is $\tbinom{L/2}{n}^2$: 
\begin{align}
j^{(2)}=\underbrace{\underbrace{000\cdots 0}_{L/2-n}\underbrace{111\cdots 1}_{n}}_{\tbinom{L/2}{n} {\rm combinations}}\underbrace{\underbrace{000\cdots 0}_{n}\underbrace{111\cdots 1}_{L/2-n}}_{\tbinom{L/2}{n} {\rm combinations}},
\end{align}
It is worth mentioning that, of course, we satisfy a condition of $\tbinom{L}{L/2}=\sum_{n=0}^{L/2}\tbinom{L/2}{n}^2$. 
Third, we change the sequence $j^{(1)}$ and for each sequence $j^{(1)}$ the number of the sequences $j^{(2)}$ is $\tbinom{L/2}{n}^2$.
Hence, the number of the sequences is
$
\tbinom{L}{L/2}\times\tbinom{L/2}{n}^2.
$
%
Finally, 
let us count the number of sets of $j^{(1)}$ and $j^{(2)}$ such that Eq.\ (B19) should be satisfied. 
This is calculated as follows: 
\begin{align}
\frac{\tbinom{L}{L/2}\times\tbinom{L/2}{n}^2}{\tbinom{L}{2n}},
\label{appeq:numberofsequencej1j2}
\end{align}
and this is summarized in Table \ref{table:numberofsequencej1j2}.

\begin{table}[htb]
\centering
\caption{the number of the sequences of $j^{(1)}+j^{(2)}$}
  \begin{tabular}{|c|c|c|c|} \hline
$n$&sequence&combination&degree of duplication\\\hline\hline
  $0$&  $\underbrace{000\cdots 0}_{L}$ &$\tbinom{L}{0}$ & $\dfrac{\tbinom{L}{L/2}\times\tbinom{L/2}{0}^2}{\tbinom{L}{0}}$\\ \hline 
  $1$&  $\underbrace{000\cdots 0}_{L-2}\underbrace{11}_{2}$ &$\tbinom{L}{2}$ & $\dfrac{\tbinom{L}{L/2}\times\tbinom{L/2}{1}^2}{\tbinom{L}{2}}$\\ \hline
  $2$&  $\underbrace{000\cdots 0}_{L-4}\underbrace{1111}_{4}$ &$\tbinom{L}{4}$ & $\dfrac{\tbinom{L}{L/2}\times\tbinom{L/2}{2}^2}{\tbinom{L}{4}}$\\ \hline
\vdots&\vdots&\vdots&\vdots\\\hline
  $L/2$ & $\underbrace{111\cdots 1}_{L}$ &$\tbinom{L}{L}$ & $\dfrac{\tbinom{L}{L/2}\times\tbinom{L/2}{L/2}^2}{\tbinom{L}{L}}$\\ \hline 
  \end{tabular}
\label{table:numberofsequencej1j2}
\end{table}
%

If we fix a sequence $j^{(1)}+j^{(2)}=\underbrace{000\cdots 0}_{L-2n}\underbrace{111\cdots 1}_{2n}$, then the sequence $j^{(3)}+j^{(4)}=\underbrace{000\cdots 0}_{L-2n}\underbrace{111\cdots 1}_{2n}$ is uniquely determined such that $j^{(1)}+j^{(2)}+j^{(3)}+j^{(4)}\equiv 00\cdots0$.
From this, we obtain
\begin{align}
\bra{D^{L}_{L/2}}_x\hat{\rho} (t)\ket{D^{L}_{L/2}}_x&=\frac{1}{2^{L}\tbinom{L}{L/2}^2}\sum_{n=0}^{L/2}\underbrace{\left(1+e^{-\left(\frac{t}{T_2}\right)^2}\right)^{L-2n}\left(1-e^{-\left(\frac{t}{T_2}\right)^2}\right)^{2n}}_{{\rm Table \,I}}\underbrace{\left(\frac{\tbinom{L}{L/2}\times\tbinom{L/2}{n}^2}{\tbinom{L}{2n}}\right)^2}_{{\rm duplication\, Eq.\, (\ref{appeq:numberofsequencej1j2})}}\times\underbrace{\tbinom{L}{2n}}_{{\rm combination}}+O((\omega_s(r_n,z_n) t)^2)\\
&=\frac{1}{2^{L}}\sum_{n=0}^{L/2}\frac{\tbinom{L/2}{n}^4}{\tbinom{L}{2n}}\left(1+e^{-\left(\frac{t}{T_2}\right)^2}\right)^{L-2n}\left(1-e^{-\left(\frac{t}{T_2}\right)^2}\right)^{2n}+O((\omega_s(r_n,z_n) t)^2)\\
&=e^{-\frac{L}{2}\left(\frac{t}{T_2}\right)^2}\sum_{n= 0}^{L/2}\frac{\tbinom{L/2}{n}^4}{\tbinom{L}{2n}}\left[\tanh{\frac{1}{2}\left(\frac{t}{T_2}\right)^2}\right]^{2n}+O((\omega_s(r_n,z_n) t)^2)+O(L^{-1}).
\label{appeq:firsttermp}
\end{align}

\subsection{Second term calculation: $\bra{D^{L}_{L/2+1}}_x\hat{\rho} (t)\ket{D^{L}_{L/2+1}}_x$}
The second term $\bra{D^{L}_{L/2+1}}_x\hat{\rho} (t)\ket{D^{L}_{L/2+1}}_x$ gives
\begin{align}
&\frac{1}{2^{2L}}\sum_{m,m'} \zeta(m)\zeta(m')\xi(m)\xi(m')\times \exp{\left[i\sum_n \frac{\omega_s(r_n,z_n) }{2}t\{(-1)^{m_n}-(-1)^{m_n'}\}\right]}\times\prod_n\left(\delta_{m_n,m_n'}+(1-\delta_{m_n,m_n'})e^{-\left(\frac{t}{T_2}\right)^2}\right)\\
&=\frac{1}{2^{L} \tbinom{L}{L/2}\tbinom{L}{L/2+1}}\sum_{j^{(1)}j^{(2)}l^{(1)}l^{(2)}}\prod_n\left(\delta_{j^{(1)}_n+j^{(2)}_n+l^{(1)}_n+l^{(2)}_n \equiv 0}+\delta_{j^{(1)}_n,l^{(1)}_n}(1+i\omega_s(r_n,z_n) t )e^{-\left(\frac{t}{T_2}\right)^2}+\delta_{j^{(2)}_n,l^{(2)}_n}(1-i\omega_s(r_n,z_n) t )e^{-\left(\frac{t}{T_2}\right)^2}-e^{-\left(\frac{t}{T_2}\right)^2}\right)\notag\\
&\qquad\qquad\qquad\qquad\qquad\qquad\qquad\qquad\qquad\qquad\qquad\qquad\qquad\qquad\qquad\qquad\qquad\qquad\qquad+O((\omega_s(r_n,z_n) t)^2).
\label{appeq:secondterm}
\end{align}
Note that Eq.\ (\ref{appeq:secondterm}) is equal to Eq.\ (\ref{appeq:firstterm}) except the range of the sum $\sum_{j^{(1)}j^{(2)}l^{(1)}l^{(2)}}$.
Therefore, we investigate the sequence $j^{(1)}+l^{(1)}$ such that $j^{(1)}+l^{(1)}+j^{(2)}+l^{(2)} \equiv 00\cdots0$.
From the same discussion as Eq.\ (\ref{appeq:numberofsequencej1j2}), the degree of duplication for each sequence is given as
\begin{align}
\frac{\tbinom{L}{L/2}\times\tbinom{L/2}{n-1}\times\tbinom{L/2}{n}}{\tbinom{L}{2n-1}},
\label{appeq:numberofsequencej1l1}
\end{align}
and this is summarized in Table \ref{table:numberofsequencej1l1}.
\begin{table}[htb]
\centering
\caption{the number of the sequences of $j^{(1)}+l^{(1)}$}
  \begin{tabular}{|c|c|c|c|} \hline
$n$&sequence&combination&degree of duplication\\\hline\hline
$1$&    $\underbrace{000\cdots 0}_{L-1}\underbrace{1}_{1}$ &$\tbinom{L}{1}$ & $\dfrac{\tbinom{L}{L/2}\times\tbinom{L/2}{0}\times \tbinom{L/2}{1}}{\tbinom{L}{1}}$\\ \hline 
 $2$&   $\underbrace{000\cdots 0}_{L-3}\underbrace{111}_{3}$ &$\tbinom{L}{3}$ & $\dfrac{\tbinom{L}{L/2}\times\tbinom{L/2}{1}\times \tbinom{L/2}{2}}{\tbinom{L}{3}}$\\ \hline
  $3$&  $\underbrace{000\cdots 0}_{L-5}\underbrace{11111}_{5}$ &$\tbinom{L}{5}$ & $\dfrac{\tbinom{L}{L/2}\times\tbinom{L/2}{2}\times \tbinom{L/2}{3}}{\tbinom{L}{5}}$\\ \hline
\vdots&\vdots&\vdots&\vdots\\\hline
 $L/2$&   $\underbrace{0}_{1}\underbrace{111\cdots 1}_{L-1}$ &$\tbinom{L}{L-1}$ & $\dfrac{\tbinom{L}{L/2}\times\tbinom{L/2}{L/2-1}\times \tbinom{L/2}{L/2}}{\tbinom{L}{L-1}}$\\ \hline 
  \end{tabular}
\label{table:numberofsequencej1l1}
\end{table}
From this, we obtain
\begin{align}
&\bra{D^{L}_{L/2+1}}_x\hat{\rho} (t)\ket{D^{L}_{L/2+1}}_x\notag\\
&=\frac{1}{2^{L}\tbinom{L}{L/2}\tbinom{L}{L/2+1}}\sum_{n=0}^{L/2}\underbrace{\left(1+e^{-\left(\frac{t}{T_2}\right)^2}\right)^{L-2n}\left(1-e^{-\left(\frac{t}{T_2}\right)^2}\right)^{2n}}_{{\rm Table \,I}}\underbrace{\left(\frac{\tbinom{L}{L/2}\times\tbinom{L/2}{n-1}\times\tbinom{L/2}{n}}{\tbinom{L}{2n-1}}\right)^2}_{{\rm duplication \, Eq.\, (\ref{appeq:numberofsequencej1l1})}}\times\underbrace{\tbinom{L}{2n-1}}_{{\rm combination}}+O((\omega_s(r_n,z_n) t)^2)\\
&=\frac{\tbinom{L}{L/2}}{2^{L} \tbinom{L}{L/2+1}}\sum_{n= 1}^{L/2}\frac{\tbinom{L/2}{n-1}^2\times\tbinom{L/2}{n}^2}{\tbinom{L}{2n-1}}\left(1+e^{-\left(\frac{t}{T_2}\right)^2}\right)^{L-2n+1}\left(1-e^{-\left(\frac{t}{T_2}\right)^2}\right)^{2n-1}+O((\omega_s(r_n,z_n) t)^2)\\
&=e^{-\frac{L}{2}\left(\frac{t}{T_2}\right)^2}\sum_{n= 1}^{L/2}\frac{\tbinom{L/2}{n-1}^2\times\tbinom{L/2}{n}^2}{\tbinom{L}{2n-1}}\left[\tanh{\frac{1}{2}\left(\frac{t}{T_2}\right)^2}\right]^{2n-1}+O((\omega_s(r_n,z_n) t)^2)+O(L^{-1}).
\label{appeq:secondtermp}
\end{align}

\subsection{Third term calculation: $\mathrm{Im}[\bra{D^{L}_{L/2}}_x\hat{\rho} (t)\ket{D^{L}_{L/2+1}}_x]$}
The third term gives
\begin{align}
&\bra{D^{L}_{L/2}}_x\hat{\rho} (t)\ket{D^{L}_{L/2+1}}_x
=\frac{1}{2^{L} \tbinom{L}{L/2}^{3/2}\tbinom{L}{L/2+1}^{1/2}}\sum_{j^{(1)}j^{(2)}j^{(3)}l^{(1)}}\prod_n\left(\delta_{j^{(1)}_n+j^{(2)}_n+j^{(3)}_n+l^{(1)}_n \equiv 0}\right.\notag\\
&\left.+\delta_{j^{(1)}_n,j^{(2)}_n}(1+i\omega_s(r_n,z_n) t )e^{-\left(\frac{t}{T_2}\right)^2}+\delta_{j^{(3)}_n,l^{(1)}_n}(1-i\omega_s(r_n,z_n) t )e^{-\left(\frac{t}{T_2}\right)^2}-e^{-\left(\frac{t}{T_2}\right)^2}\right)+O((\omega_s(r_n,z_n) t)^2).
\label{appeq:thirdterm}
\end{align}
Note that Eq.\ (\ref{appeq:thirdterm}) is also equal to Eq.\ (\ref{appeq:firstterm}) except the range of the sum $\sum_{j^{(1)}j^{(2)}j^{(3)}l^{(1)}}$.
Therefore, we investigate the sequence $j^{(1)}+j^{(2)}$ and $j^{(3)}+l^{(1)}$  such that $j^{(1)}+j^{(2)}+j^{(3)}+l^{(1)} \equiv \underbrace{00\cdots0}_{L-1}1$.
The degree of duplication for each sequence is discussed in the previous subsections.
More specifically, the number of duplications of $j^{(1)}+j^{(2)}$ is $\frac{\tbinom{L}{L/2}\times\tbinom{L/2}{n}^2}{\tbinom{L}{2n}}$ as we discussed. 
Also, the number of duplications of $j^{(3)}+l^{(1)}$ is $\frac{\tbinom{L}{L/2}\times\tbinom{L/2}{n-1}\times\tbinom{L/2}{n}}{\tbinom{L}{2n-1}}$, as we discussed. 
From this, we obtain
\begin{align}
&\bra{D^{L}_{L/2}}_x\hat{\rho} (t)\ket{D^{L}_{L/2+1}}_x\notag\\
&=\frac{\left(\frac{1}{L}\sum_j(-i\omega_s(r_j,z_j) t )e^{-\left(\frac{t}{T_2}\right)^2}\right)}{2^{L} \tbinom{L}{L/2}^{3/2}\tbinom{L}{L/2+1}^{1/2}}\sum_{n=1}^{L/2}\underbrace{\left(1+e^{-\left(\frac{t}{T_2}\right)^2}\right)^{L-2n}\left(1-e^{-\left(\frac{t}{T_2}\right)^2}\right)^{2n-1}}_{{\rm Table \,I}}\underbrace{\frac{\tbinom{L}{L/2}\times\tbinom{L/2}{n}^2}{\tbinom{L}{2n}}}_{{\rm duplication\, Eq.\ (\ref{appeq:numberofsequencej1j2})}}\times\underbrace{\frac{\tbinom{L}{L/2}\times\tbinom{L/2}{n-1}\times\tbinom{L/2}{n}}{\tbinom{L}{2n-1}}}_{{\rm duplication\, Eq.\, (\ref{appeq:numberofsequencej1l1})}}\times\underbrace{\tbinom{L}{2n-1}}_{{\rm combination}}\times(2n)\notag\\
&+\frac{\left(\frac{1}{L}\sum_j(i\omega_s(r_j,z_j) t )e^{-\left(\frac{t}{T_2}\right)^2}\right)}{2^{L} \tbinom{L}{L/2}^{3/2}\tbinom{L}{L/2+1}^{1/2}}\sum_{n=1}^{L/2}\underbrace{\left(1+e^{-\left(\frac{t}{T_2}\right)^2}\right)^{L-2n}\left(1-e^{-\left(\frac{t}{T_2}\right)^2}\right)^{2n-1}}_{{\rm Table \,I}}\underbrace{\frac{\tbinom{L}{L/2}\times\tbinom{L/2}{n-1}^2}{\tbinom{L}{2(n-1)}}}_{{\rm duplication\, Eq.\ (\ref{appeq:numberofsequencej1j2})}}\times\underbrace{\frac{\tbinom{L}{L/2}\times\tbinom{L/2}{n-1}\times\tbinom{L/2}{n}}{\tbinom{L}{2n-1}}}_{{\rm duplication\, Eq.\ (\ref{appeq:numberofsequencej1l1})}}\times\underbrace{\tbinom{L}{2n-1}}_{{\rm combination}}\times(2n-1)\notag
\end{align}
and therefore
\begin{align}
&\mathrm{Im}[\bra{D^{L}_{L/2}}_x\hat{\rho} (t)\ket{D^{L}_{L/2+1}}_x]
%
=-\frac{e^{-\frac{L}{2}\left(\frac{t}{T_2}\right)^2}}{2L}\left(\sum_j\omega_s(r_j,z_j)t\right)\sum_{n= 1}^{L/2}\frac{\tbinom{L/2}{n}^3\tbinom{L/2}{n}(2n)}{\tbinom{L}{2n-1}}\left[\tanh{\frac{1}{2}\left(\frac{t}{T_2}\right)^2}\right]^{2n-1}\notag\\
&+\frac{e^{-\frac{L}{2}\left(\frac{t}{T_2}\right)^2}}{2L}\left(\sum_j\omega_s(r_j,z_j)t\right)\sum_{n= 1}^{L/2}\frac{\tbinom{L/2}{n-1}^3\tbinom{L/2}{n}(2n-1)}{\tbinom{L}{2n-2}}\left[\tanh{\frac{1}{2}\left(\frac{t}{T_2}\right)^2}\right]^{2n-2}+O((\omega_s(r_n,z_n)t)^2)+O(L^{-1}). 
\label{appeq:thirdtermp} 
\end{align}

\subsection{Derivation of the explicit form $p$ in Eq.\ (\ref{appeq:explicitformp})}
As described in the main text, we set $t=\frac{T_2}{\sqrt{L}}u$ in Eqs.\ (\ref{appeq:firsttermp}), (\ref{appeq:secondtermp}), and (\ref{appeq:thirdtermp}):
\begin{align}
p&=
\frac{e^{-\frac{u^2}{2}}}{2}\sum_{n= 0}^{L/2}\frac{\tbinom{L/2}{n}^4}{\tbinom{L}{2n}}\left(\frac{u^2}{2L}\right)^{2n}+\frac{e^{-\frac{u^2}{2}}}{2}\sum_{n= 1}^{L/2}\frac{\tbinom{L/2}{n-1}^2\times\tbinom{L/2}{n}^2}{\tbinom{L}{2n-1}}\left(\frac{u^2}{2L}\right)^{2n-1}+\frac{ T_2}{2L\sqrt{L}}ue^{-\frac{u^2}{2}}\left(\sum_j\omega_s(r_j,z_j)\right)\sum_{n= 1}^{L/2}\frac{\tbinom{L/2}{n}^3\tbinom{L/2}{n}(2n)}{\tbinom{L}{2n-1}}\left(\frac{u^2}{2L}\right)^{2n-1}\notag\\
&-\frac{ T_2}{2L\sqrt{L}}ue^{-\frac{u^2}{2}}\left(\sum_j\omega_s(r_j,z_j)\right)\sum_{n= 1}^{L/2}\frac{\tbinom{L/2}{n-1}^3\tbinom{L/2}{n}(2n-1)}{\tbinom{L}{2n-2}}\left(\frac{u^2}{2L}\right)^{2n-2}+O((\omega_s(r_n,z_n)t)^2)+O(L^{-1}).
\end{align}
Here, we rewrite the first two terms
\begin{align}
&\frac{e^{-\frac{u^2}{2}}}{2}\sum_{n= 0}^{L/2}\frac{\tbinom{L/2}{n}^4}{\tbinom{L}{2n}}\left(\frac{u^2}{2L}\right)^{2n}+\frac{e^{-\frac{u^2}{2}}}{2}\sum_{n= 1}^{L/2}\frac{\tbinom{L/2}{n-1}^2\times\tbinom{L/2}{n}^2}{\tbinom{L}{2n-1}}\left(\frac{u^2}{2L}\right)^{2n-1}\\
&=\frac{e^{-\frac{u^2}{2}}}{2}\sum_{n= 0}^{\infty}\frac{(2n)!}{(n!)^4}\left(\frac{u^2}{8}\right)^{2n}+\frac{e^{-\frac{u^2}{2}}}{2}\sum_{n= 1}^{\infty}\frac{(2n-1)!}{(n!)^2((n-1)!)^2}\left(\frac{u^2}{8}\right)^{2n-1}+O(L^{-1})\\
&=\frac{e^{-\frac{u^2}{2}}}{2}I_{0}(u^2/4)\left[I_{0}(u^2/4)+I_{1}(u^2/4)\right],
\end{align}
where $I_{\alpha}(x)$ is the modified Bessel function
$
I_{\alpha}(x)=\sum_{m=0}^\infty \frac{1}{m!\Gamma(m+\alpha+1)}\left(\frac{x}{2}\right)^{2m+\alpha},
$
and $\Gamma(x)$ is the Gamma function.
%
Also, we also rewrite 
\begin{align}
&\sum_{n= 1}^{L/2}\left(\frac{\tbinom{L/2}{n}^3\tbinom{L/2}{n}(2n)}{\tbinom{L}{2n-1}}\left(\frac{u^2}{2L}\right)^{2n-1}-\frac{\tbinom{L/2}{n-1}^3\tbinom{L/2}{n}(2n-1)}{\tbinom{L}{2n-2}}\left(\frac{u^2}{2L}\right)^{2n-2}\right)\\
&=\frac{L}{2}\left(I_{0}(u^2/4)-I_{1}(u^2/4)\right)^2+O(1),
\end{align}
and consequently we obtain $p$ in Eq.\ (\ref{appeq:explicitformp}).

\section{Negligibility of interaction between probe spins}
In this section, we show that the interaction between probe spins can be neglected effectively.
We consider that the probe spins are an ensemble of NV centers and the target spin is qubit.
It is known that an application of electric fields can suppress
magnetic interaction of the NV centers \cite{dolde2011electric,iwasaki2017direct},
and we use these experimental facts for the suppression of the dipole-dipole interaction
between the NV centers.
NV centers are regarded as spin 1 with three levels $\ket{0},\ket{+1},\ket{-1}$, and spin operators $\hat{S}_x,\hat{S}_y,\hat{S}_z$ are defined as follows:
\begin{align}
\hat{S}_x&=\ket{B}\bra{0}+\ket{0}\bra{B}\\
\hat{S}_y&=-i\ket{D}\bra{0}+i\ket{0}\bra{D}\\
\hat{S}_z&=\ket{B}\bra{D}+\ket{D}\bra{B},
\end{align}
where $\ket{B}=(\ket{+1}+\ket{-1})/\sqrt{2}$ and $\ket{D}=(\ket{+1}-\ket{-1})/\sqrt{2}$ are a bright state and a dark state, respectively.
We consider an ensemble of NV centers coupled with a single spin and that Hamiltonian is given by \cite{bauch2019decoherence,hayashi2020experimental,matsuzaki2016optically,kubo2011hybrid,zhu2014observation}
\begin{align}
 H_{{\rm NV-NV}}=&\sum_j\left[D_0 \hat{S}_{z,j}^2 +E(\hat{S}_{x,j}^2 -\hat{S}_{y,j}^2)
+ g^{(1)}_{j}(\hat{S}_{x,j} \hat{\sigma }_{x}^{({\rm T})}+ \hat{S}_{y,j} \hat{\sigma }_{y}^{({\rm T})})
+ g^{(2)}_{j}\hat{S}_{z,j} \hat{\sigma }_{z}^{({\rm T})}\right] \notag\\
&+\sum_{j,k}\left[g^{(1)}_{j,k}\left(\hat{S}_{x,j}\hat{S}_{x,k}+\hat{S}_{y,j}\hat{S}_{y,k}\right)+g^{(2)}_{j,k}\hat{S}_{z,j}\hat{S}_{z,k}\right]+\frac{\omega^{({\rm T})} }{2}\hat{\sigma }_{z}^{({\rm T})}+\Omega(t)\cos{(\omega^{({\rm T})} t)}\hat{\sigma }_{x}^{({\rm T})},
\end{align}
where $D_0$ is the zero-field splitting term and $E$ denotes the electric fields  and $\hat{\sigma }_{x}^{({\rm T})}$, $\hat{\sigma }_{y}^{({\rm T})}$, and $\hat{\sigma }_{z}^{({\rm T})}$ are the Pauli matrices of the target spin, which are the same as those in Eq.~(3).
Here, $g^{(1)}_{j}$ and $g^{(2)}_{j}$ denote the dipole-dipole interaction between each probe spin and the target spin, and $g^{(1)}_{j,k}$ and $g^{(2)}_{j,k}$ represent the dipole-dipole interaction between NV centers, and the last term represents the dynamical decoupling \cite{schmitt2017submillihertz}, and $\Omega(t)$ is a set of sharp $\pi$ pulses at regular intervals $\pi /(2E)$.
Let us go to the rotating frame defined by $H_0=D_0 \sum_j\hat{S}_{z,j}^2 +\frac{\omega^{({\rm T})} }{2}\hat{\sigma }_z^{({\rm T})}$.
Under the rotating wave approximation assuming $\Omega(t)\ll \omega^{({\rm T})}$, we obtain
\begin{align}
H_{{\rm NV-NV}} \simeq &\sum_j\left[ E(|B\rangle_j \langle B|_j  -|D\rangle_j \langle D|_j)
+ g^{(2)}_{j}\hat{S}_{z,j}\hat{\sigma }_z^{({\rm T})}\right] +\sum_{j,k}\left[g^{(1)}_{j,k}\left(\hat{S}_{x,j}\hat{S}_{x,k}+\hat{S}_{y,j}\hat{S}_{y,k}\right)+g^{(2)}_{j,k}\hat{S}_{z,j}\hat{S}_{z,k}\right]+\frac{\Omega(t)}{2}\hat{\sigma}_{x}^{({\rm T})}\\
\simeq &\sum_j\left[ E(|B\rangle_j \langle B|_j  -|D\rangle_j \langle D|_j)
+ h(t+\pi/(4E))g^{(2)}_{j}\hat{S}_{z,j}
\hat{\sigma }_z^{({\rm T})}\right]+\sum_{j,k}\left[g^{(1)}_{j,k}\left(\hat{S}_{x,j}\hat{S}_{x,k}+\hat{S}_{y,j}\hat{S}_{y,k}\right)+g^{(2)}_{j,k}\hat{S}_{z,j}\hat{S}_{z,k}\right],
\end{align}
where $h(t)$ is a square function: $h(t)=1$ ($(2n-2)\pi \le 2Et \le (2n-1)\pi$), while $h(t)=-1$ ($(2n-1)\pi \le 2Et \le 2n\pi$), and $n$ is an arbitrary natural number.
It is worth mentioning that $h(t)$ can be rewritten as
\begin{align}
h(t)=\frac{4}{\pi}\sum_{n{\rm ~:~odd}} \frac{1}{n}\sin{(2Ent)}.
\end{align}
So we have 
\begin{align}
h(t+\pi/(4E))=\frac{4}{\pi}\sum_{n{\rm ~:~odd}} \frac{1}{n}\sin{\left(2Ent+\frac{\pi}{2}n\right)}.
\end{align}
In the interaction picture defined by $H_0=\sum_jE\left( |B\rangle_j \langle B|_j  -|D\rangle_j \langle D|_j\right)$ and under the rotating wave approximation, we obtain
\begin{align}
 H_{{\rm NV-NV}}\simeq& \frac{2}{\pi} \sum_jg^{(2)}_{j}\hat{S}_{z,j} \hat{\sigma }_z+\sum_{j,k}\left[g^{(1)}_{j,k}\left(|B0\rangle \langle 0B|  +|0B\rangle \langle B0|+|D0\rangle \langle 0D|  +|0D\rangle \langle D0|\right)+g^{(2)}_{j,k}\left(|BD\rangle \langle BD|  +|DB\rangle \langle DB|\right)\right].
\end{align}
The dynamical decoupling has been realized with the electron spins
\cite{de2010universal}.
The dynamical decoupling with the electron spins
has been used to detect nuclear spins
\cite{taminiau2012detection}. 
Moreover, the theoretical treatment of the dynamical decoupling 
was introduced in \cite{chin2012quantum,schmitt2017submillihertz}, which we adopt in this paper.
Here, we define the interaction Hamiltonian $H'_{{\rm NV-NV}},H''_{{\rm NV-NV}}$ by
\begin{align}
H'_{{\rm NV-NV}}&=\sum_{j,k}g^{(1)}_{j,k}\left(|B0\rangle \langle 0B|  +|0B\rangle \langle B0|+|D0\rangle \langle 0D|  +|0D\rangle \langle D0|\right)\\
H''_{{\rm NV-NV}}&=\sum_{j,k}g^{(2)}_{j,k}\left(|BD\rangle \langle BD|  +|DB\rangle \langle DB| \right).
\end{align}
These interaction Hamiltonians do not disturb the initial state $\ket{D^L_{L/2}}$
\begin{align}
\ket{D^L_{L/2}}_x=\tbinom{L}{L/2}^{-1/2}\sum_{{\rm perm}}(\ket{\underbrace{BB\cdots B}_{L/2}\underbrace{DD\cdots D}_{L/2}}),
\end{align}
This is because
\begin{align}
H'_{{\rm NV-NV}}\ket{D^L_{L/2}}=0
\end{align}
and 
\begin{align}
H''_{{\rm NV-NV}}\ket{D^L_{L/2}}=\ket{D^L_{L/2}}.
\end{align}
Here, we assume the translational invariance of the NV centers in the latter equation.
Therefore, as long as we use the Dicke state for the probe spins, the dipole-dipole interaction
does not affect the dynamics.
\end{widetext}
\bibliographystyle{apsrev4-1}

%

\end{document}